\documentclass[10pt,twocolumn,english,prd,superscriptaddress,nofootinbib,preprintnumbers,showpacs,floatfix]{revtex4-2}

\usepackage{mathrsfs}
\usepackage{yfonts}
\usepackage{tipa}
\usepackage{bm}
\usepackage{bbm}
\usepackage{amsmath}
\usepackage{amssymb}
\usepackage{amsthm}
\usepackage{amsfonts}
\usepackage{mathtools}
\usepackage{latexsym}
\usepackage{relsize}
\usepackage[english]{babel}
\usepackage{booktabs}
\usepackage[iso-8859-7]{inputenc}
\usepackage{tikz}
\usepackage{array}

\newcommand{\bea}{\begin{eqnarray}}
\newcommand{\eea}{\end{eqnarray}}
\newcommand{\be}{\begin{equation}}
\newcommand{\ee}{\end{equation}}
\newcommand{\ba}{\begin{align}}
\newcommand{\ea}{\end{align}}

\begin{document}

\title{Exact solutions, trajectories and radiation patterns in the classical relativistic St\"{o}rmer
problem}
\author{Tiberiu Harko}
\email{tiberiu.harko@aira.astro.ro}
\affiliation{Department of Physics, Babe\c s-Bolyai University, Kog\u alniceanu Street,
Cluj-Napoca 400084, Romania,} 
\affiliation{Astronomical Observatory, 19
Cire\c silor Street, 400487 Cluj-Napoca, Romania}
\author{Francisco S. N. Lobo}
\email{fslobo@ciencias.ulisboa.pt}
\affiliation{Instituto de Astrof\'{i}sica e Ci\^{e}ncias do Espa\c{c}o, Faculdade de Ci\^{e}ncias da Universidade de Lisboa, Edificio C8, Campo Grande, P-1749-016 Lisbon, Portugal }
\affiliation{Departamento de F\'{i}sica, Faculdade de Ci\^{e}ncias da Universidade de Lisboa, Edif\'{i}cio C8, Campo Grande, P-1749-016 Lisbon, Portugal}

\begin{abstract}
We investigate the relativistic generalization of the classical St\"{o}rmer problem, which describes the motion of charged particles in a purely magnetic dipole field. By incorporating special relativistic effects, the particle dynamics is governed by a strongly nonlinear system of second-order differential equations derived from the Lorentz force law. We present a rigorous and fully covariant derivation of the relativistic equations of motion, together with the associated conservation laws.
An exact solution for planar motions is obtained in parametric form, providing analytical insight into the structure of the trajectories. In addition, we perform a detailed numerical analysis of the particle dynamics across both nonrelativistic and relativistic regimes, exploring a range of initial conditions and highlighting the impact of relativistic corrections.
The electromagnetic radiation emitted by the accelerated charges is also examined. We compute the time dependence of the total radiated power and determine the corresponding frequency spectrum. Our results provide a comprehensive characterization of magnetic dipole--type radiation associated with St\"{o}rmer-like motion. In particular, the power spectral density consistently exhibits distinct peaks, indicating the presence of preferred frequency bands in the emitted radiation.
\end{abstract}

\maketitle

\tableofcontents

\section{Introduction}

The St\"{o}rmer problem \cite{St1,St2,St3,St4,St4a,St5,St6,St7}, a classic and extensively studied topic in theoretical mechanics and electromagnetism, requires solving the equation of motion of a non-relativistic charged particle in a dipole magnetic field and has important geophysical and astrophysical applications. Originally motivated by the search for a physical description and explanation of the Northern Lights (Aurora Borealis), a bright visual phenomenon caused by the interaction of charged solar wind particles with the Earth's magnetic field, the theoretical framework initially developed by St\"{o}rmer \cite{St1,St2,St3,St4} and the results it yielded have found important applications in geophysics, astrophysics, and astronomy. Indeed, the St\"{o}rmer model successfully accounts for many astrophysical and geophysical effects, including the dynamical evolution of electrons and ions within planetary radiation belts \cite{B1,B2,B3}.

The classical St\"{o}rmer problem (CSP) is formulated for a non-relativistic particle of mass $m$, charge $Q$, and momentum $\vec{P}$ moving in a static dipole magnetic field with magnetic moment $\vec{M}$. The vector potential is given by $\vec{A}=(\vec{M}\times\vec{r})/r^{3}$, where $r=\sqrt{x^{2}+y^{2}+z^{2}}$ \cite{B2}, and the Hamiltonian takes the standard minimal-coupling form \cite{Jack,LL}
\begin{equation}
	H=\frac{1}{2m}\left(\vec{P}-\frac{Q}{c}\vec{A}\right)^{2}.
\end{equation}
Aligning the magnetic moment with the $z$-axis, $\vec{M}=(0,0,M_{z})$, setting the mass to unity $m=1$, and introducing the effective coupling constant $\alpha = QM_{z}/c$, the Hamiltonian reduces to
\begin{equation}
	H=\frac{1}{2}\left[ \left( P_{x}+\frac{\alpha y}{r^{3}}\right) ^{2} + \left( P_{y}-\frac{\alpha x}{r^{3}}\right) ^{2} + P_{z}^{2} \right],
\end{equation}
which can be expanded as
\begin{equation}
	H=\frac{1}{2}\left( P_{x}^{2}+P_{y}^{2}+P_{z}^{2}\right) +\frac{\alpha}{r^{3}}\left( yP_{x}-xP_{y}\right) +\frac{\alpha^{2}}{2r^{6}}\left( x^{2}+y^{2}\right). \label{H1}
\end{equation}

Solving the equations of motion derived from the Hamiltonian (\ref{H1}) and investigating the physical properties of their solutions constitutes the core of the Classical St\"{o}rmer Problem (CSP). The system admits two constants of motion: the Hamiltonian (\ref{H1}) itself and the projection of the angular momentum onto the direction of the magnetic moment, $L_{z}=xP_{y}-yP_{x}$ \cite{B1,B2}. By means of the Ziglin-Yoshida method, the CSP has been proven to be non-integrable \cite{Int}. A pivotal result emerging from the study of the CSP is the existence of allowed and forbidden regions for charged particles in a dipole magnetic field, as well as the existence of particle storage regions. The analysis reveals that charged particles are either trapped within the Earth's magnetosphere or escape to infinity; the trapping region is bounded by a torus-like surface, which in the case of the Earth corresponds to the Van Allen inner radiation belt. Inside this trapping region, the particle dynamics can be periodic, quasiperiodic, or chaotic \cite{Dilao}. These results are essential for understanding the physical properties of the Earth's magnetic field and the characteristics of the Van Allen radiation belts \cite{VA1,VA2}.

The Classical St\"{o}rmer problem has been extended and generalized to many other physical configurations, with important applications in geophysics, astronomy, and astrophysics. The Classical Rotational St\"{o}rmer Problem (CRoSP) studies the motion of a charged particle in the field of a rotating, uniformly magnetized celestial body; the trapping of a particle in the electromagnetic field of a parallel rotator was investigated in \cite{Schust}. For a particle moving in an electric field with potential $A_0=A_0(R,z)$ in the presence of the dipole magnetic field, the Lagrangian is given by
\begin{equation}
	L = -mc^2\sqrt{1-\frac{v^2}{c^2}} + \frac{e\mu}{c}\frac{R^2\dot{\phi}}{(R^2+z^2)^{3/2}} - eA_0,
\end{equation}
and in this configuration two or even three disconnected torus-shaped trapping regions may exist.

When the combined effects of the rotationally induced electric field and gravitation are taken into account, one obtains the Classical Rotational Gravitational St\"{o}rmer Problem (CRoGSP), which was studied in the presence of both fields in \cite{How} and \cite{Dull}. In cylindrical coordinates $(\rho,\phi,z)$, the inertial-frame Hamiltonian reads \cite{How}
\begin{equation}
	H = \frac{1}{2m}\left(p_{\rho}^2 + p_z^2\right) + \frac{1}{2m\rho^2}\left(p_{\phi} - \frac{q}{c}\Psi\right)^2 + U + \frac{q\Omega}{c}\Psi,
\end{equation}
where $U$ is the gravitational potential, the stream function is $\Psi = (x^2+y^2)/r^3$, and the electric field is described by $q\vec{E} = -\gamma\Omega\nabla\Psi$ with $\gamma = qM/c$ \cite{Dull}. The inclusion of the gravitational force leads to stable circular orbits that reside in a plane located above (or below) the equatorial plane of the central massive object \cite{Dull}.

The motion of a charged particle around a rotating magnetic body was considered in \cite{In}, where the physical model consists of a magnetic dipole field together with a corotating electric field. The Hamiltonian of this St\"{o}rmer problem is given by
\bea
H&=&\frac{1}{2m}\left(p_x^2+p_y^2+p_z^2\right)-\frac{GM_pm}{r}-\frac{\mu q}{mc}\frac{L_z}{r^3}\nonumber\\
&&+\frac{\mu ^2 q^2}{2mc^2}\frac{x^2+y^2}{r^6}+\frac{q\mu \omega }{c}\Psi.
\eea
The corresponding dynamical system was studied in the most reduced phase space: all equilibrium points were located and their stability was investigated. The role played by the oblateness of the central object was addressed in \cite{In1}, where the Hamiltonian takes the form
\bea
H&=&\frac{1}{2}\left(p_{\rho}^2+p_z^2+\frac{p_{\phi}^2}{\rho ^2}\right)-\frac{1}{r}-\delta \frac{p_{\phi}}{r^3}+\frac{\delta ^2}{2}\frac{\rho ^2}{r^6}+\delta \beta \frac{\rho ^2}{r^3}\nonumber\\
&&+3J_2\frac{z^2}{2r^5}-\frac{J_2}{2r^3},
\eea
with $r = \sqrt{\rho^2 + z^2}$. Here $J_2$ is a dimensionless parameter that is negative for a prolate planet and positive for an oblate one; $\delta = \omega_c/\Omega_k$, where $\Omega_k = \sqrt{M/R^3}$ is the Keplerian frequency, $\Omega_c$ is the cyclotron frequency, and $\beta = \Omega/\Omega_k$.

The motion of a relativistic charged particle in the electromagnetic field of a rotating magnetized planet with a magnetic axis inclined relative to the rotation axis was investigated in \cite{Epp0}, while the effective potential energy of particles in the field of a rotating uniformly magnetized planet was obtained in \cite{Epp}, where the electromagnetic field is represented by the superposition of a dipole magnetic field and a quadrupole electric field. Further investigations of the Classical St\"{o}rmer Problem and of charged relativistic particle motion in magnetic fields can be found in \cite{Hal, Mark, Ozturk, Pina, Kol, Leg, Ersh, Asadi}.

A semi-analytical solution to the St\"{o}rmer problem was obtained in \cite{Ersh1} by examining the motion of charged particles near the Earth's equatorial plane, where the dipole magnetic field is modeled as that of a magnetized infinite cylinder; the resulting St\"{o}rmer problem reduces to three first-order nonlinear ordinary differential equations, which were solved in polar coordinates. A distinct class of St\"{o}rmer-type problems, in which a charged particle moves in a dipolar magnetic field under the additional influence of dissipative and stochastic forces, was introduced in \cite{Moc} and termed the Stochastic-Dissipative St\"{o}rmer Problem (SDSP).

Although the trajectories of particles in St\"{o}rmer-type problems have been extensively studied, especially at a qualitative level, the electromagnetic radiation emitted by a charged particle moving in a dipole magnetic field has received considerably less attention. The radiation from ultrarelativistic electrons spiralling in the strong dipole magnetic field outside a neutron star was investigated in \cite{Th}, where the intensity and polarization of the synchrotron component were determined. Radiative properties of charged particles in dipole magnetic fields were also explored in \cite{Pap, Bur, Bur1}.

The goal of the present paper is to perform a systematic analysis of the Classical Relativistic St\"{o}rmer Problem (CRSP), the motion of a relativistic charged particle in a dipole magnetic field, by investigating its trajectories, emitted electromagnetic power, and radiation spectrum. We begin by deriving the equation of motion from a fully covariant Lagrangian formalism. Although the result can be cast in a compact covariant form, analytical and numerical investigations benefit from a component formulation, which reduces the CRSP to a system of three nonlinear ordinary differential equations. Throughout the motion the particle's energy is conserved, implying a constant Lorentz factor $\gamma$.

Obtaining exact analytical solutions of the CRSP is notoriously difficult; however, for the two-dimensional motion the system admits a simple perfectly circular orbit that reproduces the equations of a harmonic oscillator. This solution can be generalized by allowing a time-varying amplitude and frequency, which yields an exact parametric solution of the CRSP that is well suited for studying two-dimensional particle behaviour. In addition, we carry out a detailed numerical investigation of particle trajectories in both the CSP and the CRSP, exploring how variations in $\gamma$ and the initial conditions influence the trajectory and the electromagnetic emission. These parameter variations produce a rich variety of dynamical behaviours, ranging from stable, highly regular periodic orbits to chaotic motion, and are reflected in the corresponding electromagnetic spectrum.

The paper is organized as follows. In Section~\ref{sect1} we derive the relativistic equation of motion for a charged particle in a dipole magnetic field, starting from a fully covariant Lagrangian formalism and reducing the dynamics to a system of three nonlinear ordinary differential equations. Exact parametric solutions of the CRSP, valid for two-dimensional motion, are obtained in Section~\ref{sect2}. Section~\ref{sect3} describes the results of our numerical investigations of particle trajectories and electromagnetic emission in both the CSP and the CRSP. We conclude with a discussion and summary of the main results in Section~\ref{sect4}.

\section{Equations of motion of the Classical Relativistic St\"{o}rmer Problem}\label{sect1}

In this section we first write down the magnetic field configuration and the Lagrangian of the Classical Relativistic St\"{o}rmer Problem (CRSP). We then present a fully covariant derivation of the equation of motion of a relativistic charged particle in a dipole magnetic field. Finally, we discuss the electromagnetic emission properties, including the radiated power and the spectral decomposition of the radiation.

\subsection{Magnetic field, action and Lagrangian}

We adopt the metric and sign conventions of Landau--Lifshitz and Jackson. On the spacetime manifold we introduce coordinates
$x^{\mu} = (ct = x^{0},\; x = x^{1},\; y = x^{2},\; z = x^{3}) = (x^{0}, x^{i})$, for $i = 1,2,3$,
and endow the manifold with the Minkowski metric
\begin{align}
	ds^{2} &= c^{2}dt^{2} - dx^{2} - dy^{2} - dz^{2} \nonumber\\
	&= \eta_{\mu\nu}\, dx^{\mu} dx^{\nu} 
	= \eta_{00} (dx^{0})^{2} - \eta_{ij}\, dx^{i} dx^{j},
\end{align}
where the special-relativistic metric tensor has components $\eta_{00}=1$ and $\eta_{ij} = \operatorname{diag}(1,1,1)$, $i,j = 1,2,3$.

The four-velocity of the particle is defined as
\begin{equation}
	u^{\mu} = \frac{dx^{\mu}}{ds} = (u^{0}, u^{i}), \qquad i = 1,2,3,
\end{equation}
and satisfies the normalization condition $u_{\mu}u^{\mu} = 1$. The ordinary three-velocities $v^{i} = dx^{i}/dt$ are related to the four-velocity by
\begin{equation}
	u^{i} = \frac{dx^{i}}{ds} = \frac{dx^{i}}{dt}\frac{dt}{ds} = v^{i}\,\frac{dt}{ds}, \qquad i = 1,2,3,
	\label{u-v}
\end{equation}
and the squared speed is $v^{2} = \eta_{ij} v^{i} v^{j}$.

We assume that the electromagnetic potential has only a magnetic part, so $A^{\mu} = (0,\vec{A}) = (0, A^{i})$ and $A_{\mu} = (0, -A_{i})$. The magnetic field is taken to be a pure dipole with the magnetic moment $\vec{M}$ aligned along the $z$-axis, $\vec{M} = (0,0,M)$. The vector potential then reads
\begin{equation}
	\vec{A} = \frac{1}{r^{3}}\,\vec{M}\times\vec{r} = \frac{M}{r^{3}}\,(-y\,\vec{i} + x\,\vec{j}),
\end{equation}
with components
\begin{eqnarray}
	A^{i} &=& \frac{M}{r^{3}}\,(-y,\,x,\,0),
		\nonumber \\
	A_{i} & =& \frac{M}{r^{3}}\,(y,\,-x,\,0) = \frac{M}{r^{3}}\,(x^{2}, -x^{1}, 0),
		\nonumber
\end{eqnarray}
where $r = \sqrt{x^{2}+y^{2}+z^{2}}$.

The action for a charged particle in an external electromagnetic field is
\begin{align}
	S &= \int_{a}^{b} \Bigl( -mc\,ds - \frac{e}{c} A_{\mu} u^{\mu} \Bigr) \nonumber\\
	&= \int_{a}^{b} \Bigl( -mc\sqrt{u_{\mu}u^{\mu}} - \frac{e}{c} A_{\mu} u^{\mu} \Bigr) ds,
\end{align}
where $e$ and $m$ denote the charge and mass of the particle. The equations of motion follow from the Lagrangian
\begin{equation}
	L = -mc\sqrt{u_{\mu}u^{\mu}} - \frac{e}{c} A_{\mu} u^{\mu},
\end{equation}
via the Euler-Lagrange equations
\begin{equation}
	\frac{d}{ds}\frac{\partial L}{\partial u^{\mu}} - \frac{\partial L}{\partial x^{\mu}} = 0, \qquad \mu = 0,1,2,3.
\end{equation}

For a dipole magnetic field the Lagrangian becomes
\begin{eqnarray}
	L &=& -mc\sqrt{(u^{0})^{2} - (u^{1})^{2} - (u^{2})^{2} - (u^{3})^{2}}
		\nonumber\\
	&& \qquad - \frac{e}{c}\frac{M}{r^{3}}\bigl( x^{2} u^{1} - x^{1} u^{2} \bigr) \nonumber\\
	&=& -mc\sqrt{\eta_{00} u^{0} u^{0} - \eta_{jk} u^{j} u^{k}}
	+ \frac{eM}{c}\,\frac{\varepsilon_{ab}\, x^{a} u^{b}}{r^{3}}, 
	\label{Lagr}
\end{eqnarray}
for $a,b = 1,2$, where we have introduced the two-dimensional Levi-Civita symbol $\varepsilon_{ab}$ with the properties $\varepsilon_{11} = \varepsilon_{22} = 0$, $\varepsilon_{12} = -\varepsilon_{21} = 1$, and $\varepsilon_{a\ell} = 0$ for $\ell > 2$.

\subsection{The equations of motion and radiative power loss}

\subsubsection{Equations of motion}

The relativistic dynamics of a charged particle in a prescribed electromagnetic field are elegantly encoded in the Euler-Lagrange equations derived from the action principle.  For the Lagrangian (\ref{Lagr}) the variation with respect to the time component $u^{0}$ immediately yields a first integral
\begin{equation}
	-\frac{1}{2}mc\,u^{0} = -\frac{1}{2}mc^{2}\frac{dt}{ds} = -\frac{1}{2}\frac{E}{c} = \mathrm{constant},
\end{equation}
which reflects the invariance of the action under time translations. In three-dimensional language this translates into the conservation of the relativistic energy
\begin{eqnarray*}
	E &=& \gamma mc^{2} = \frac{mc^{2}}{\sqrt{1-v^{2}/c^{2}}} = \mathrm{constant}, \\[2pt]
	\frac{dt}{ds} &=& \frac{E}{mc^{3}} = \frac{\gamma}{c},
\end{eqnarray*}
so that the Lorentz factor $\gamma$ remains a constant of the motion.

The spatial components $i=1,2,3$ encode the genuine Lorentz force law. Computing the partial derivatives of $L$ with respect to the four-velocity and the coordinates gives
\begin{equation}
	\frac{\partial L}{\partial u^{l}} = mc\,\eta_{kl}\,\frac{u^{k}}{\sqrt{u_{\mu}u^{\mu}}} + \frac{eM}{c}\,\varepsilon_{al}\,\frac{x^{a}}{r^{3}},
\end{equation}
and
\begin{equation}
	\frac{\partial L}{\partial x^{l}} = \frac{eM}{c}\,\frac{\varepsilon_{lb}u^{b}}{r^{3}} - \frac{3eM}{c}\,\varepsilon_{ab}\,\eta_{kl}\,\frac{x^{a}x^{k}}{r^{5}}\,u^{b},
\end{equation}
where we have made use of the radial derivative
\begin{equation}
	\frac{\partial}{\partial x^{l}}\frac{1}{r^{3}} = \frac{\partial}{\partial x^{l}}\bigl(\eta_{jk}x^{j}x^{k}\bigr)^{-3/2} = -\frac{3\eta_{kl}x^{k}}{r^{5}}.
\end{equation}
Inserting these expressions into the Euler-Lagrange equations yields the relativistic equation of motion in covariant form
\begin{eqnarray}
	\frac{d}{ds}\!\left(mc\,\eta_{kl}u^{k} + \frac{eM}{c}\,\varepsilon_{al}\,\frac{x^{a}}{r^{3}}\right)
	= \frac{eM}{c}\,\frac{\varepsilon_{la}u^{a}}{r^{3}} 
		\nonumber\\
	- \frac{3eM}{c}\,\frac{\varepsilon_{ab}x^{a}u^{b}}{r^{5}}\,\eta_{kl}x^{k}.
\end{eqnarray}
Expanding the derivative on the left-hand side and regrouping the terms we obtain a more transparent structure
\begin{eqnarray}
	mc\,\eta_{kl}\,\frac{du^{k}}{ds} &=& -2\frac{eM}{c}\,\varepsilon_{al}\,\frac{u^{a}}{r^{3}}
	- \frac{eM}{c}\,\varepsilon_{al}x^{a}\frac{d}{ds}\!\left(\frac{1}{r^{3}}\right) \nonumber\\
	&& - \frac{3eM}{c}\,\varepsilon_{ab}\,\eta_{kl}\,\frac{x^{k}x^{a}u^{b}}{r^{5}}.
\end{eqnarray}
The time derivative of $1/r^{3}$ is easily re-expressed using the chain rule
\begin{equation}
	\frac{d}{ds}\frac{1}{r^{3}} = \left(\frac{d}{dx^{j}}\frac{1}{r^{3}}\right)\frac{dx^{j}}{ds} = -\frac{3\eta_{jk}x^{k}u^{j}}{r^{5}},
\end{equation}
leading to the fully covariant equation
\begin{eqnarray}
	&& mc\,\eta_{kl}\,\frac{du^{k}}{ds} = -2\frac{eM}{c}\,\varepsilon_{al}\,\frac{u^{a}}{r^{3}} \nonumber\\
	&& \qquad + \frac{3eM}{c}\,\frac{1}{r^{5}}\Bigl(\varepsilon_{al}x^{a}\eta_{jk}x^{k}u^{j} - \eta_{kl}x^{k}\varepsilon_{ab}x^{a}u^{b}\Bigr).
\end{eqnarray}

For practical computations it is often more convenient to work with the three-velocity.  Using the relations $u^{k} = v^{k}\,dt/ds$ and $du^{k}/ds = (dv^{k}/dt)(dt/ds)^{2}$, together with $dt/ds = \gamma/c$, we obtain the equation of motion in ordinary three-dimensional notation
\begin{eqnarray}
	&&\eta_{kl}\,\frac{dv^{k}}{dt} = -2\frac{eM}{mc}\,\frac{1}{\gamma}\,\varepsilon_{al}\,\frac{v^{a}}{r^{3}} \nonumber\\
	&& \qquad + \frac{3eM}{mc}\,\frac{1}{\gamma}\,\frac{1}{r^{5}}\Bigl(\varepsilon_{al}x^{a}\eta_{jk}x^{k}v^{j} - \eta_{kl}x^{k}\varepsilon_{ab}x^{a}v^{b}\Bigr).
\end{eqnarray}
Writing the components explicitly, we obtain three coupled second-order differential equations
\begin{eqnarray}
	\frac{d^{2}x}{dt^{2}} &=& -\frac{eM}{mc}\frac{1}{\gamma}\frac{\dot{y}}{r^{3}} + \frac{3eM}{mc}\frac{1}{\gamma}\frac{z}{r^{5}}\bigl(z\dot{y} - y\dot{z}\bigr), \label{LL1}\\
	\frac{d^{2}y}{dt^{2}} &=& \frac{eM}{mc}\frac{1}{\gamma}\frac{\dot{x}}{r^{3}} - 3\frac{eM}{mc}\frac{1}{\gamma}\frac{z}{r^{5}}\bigl(\dot{x}z - \dot{z}x\bigr), \label{LL2}\\
	\frac{d^{2}z}{dt^{2}} &=& 3\frac{eM}{mc}\frac{1}{\gamma}\frac{z}{r^{5}}\bigl(\dot{x}y - \dot{y}x\bigr), \label{LL3}
\end{eqnarray}
where the dot denotes differentiation with respect to the coordinate time $t$.

To bring out the essential dynamical scales we introduce dimensionless variables by measuring lengths in units of a characteristic scale $R_{0}$ and time in units of the characteristic gyration time
\begin{equation}
	x = R_{0}X,\quad y = R_{0}Y,\quad z = R_{0}Z,\quad \tau = \frac{eM}{mcR_{0}^{3}}\,t,
\end{equation}
so that $r = R_{0}R$ with $R = \sqrt{X^{2}+Y^{2}+Z^{2}}$.  In these variables the equations of motion become
\begin{eqnarray}
	\frac{d^{2}X}{d\tau^{2}} &=& -\frac{1}{\gamma}\frac{1}{R^{3}}\frac{dY}{d\tau} + \frac{3}{\gamma}\frac{Z}{R^{5}}\left(Z\frac{dY}{d\tau} - Y\frac{dZ}{d\tau}\right), \label{F1}\\
	\frac{d^{2}Y}{d\tau^{2}} &=& \frac{1}{\gamma}\frac{1}{R^{3}}\frac{dX}{d\tau} - \frac{3}{\gamma}\frac{Z}{R^{5}}\left(Z\frac{dX}{d\tau} - X\frac{dZ}{d\tau}\right), \label{F2}\\
	\frac{d^{2}Z}{d\tau^{2}} &=& \frac{3}{\gamma}\frac{Z}{R^{5}}\left(Y\frac{dX}{d\tau} - X\frac{dY}{d\tau}\right). \label{F3}
\end{eqnarray}

The dimensionless three-velocity squared of the particle reads
\begin{eqnarray}
	v^{2} &=& \left(\frac{eM}{mcR_{0}^{2}}\right)^{2}\!\left[\left(\frac{dX}{d\tau}\right)^{2} + \left(\frac{dY}{d\tau}\right)^{2} + \left(\frac{dZ}{d\tau}\right)^{2}\right] \nonumber\\
	&=& \left(\frac{eM}{mcR_{0}^{2}}\right)^{2}\Bigg(V_X^{2}+V_Y^{2}+V_Z^{2}\Bigg) \nonumber\\
	&=& \left(\frac{eM}{mcR_{0}^{2}}\right)^{2} V^{2}.
\end{eqnarray}
Correspondingly, the Lorentz factor can be expressed as
\begin{equation}
	\gamma = \frac{1}{\sqrt{1 - \displaystyle\left(\frac{eM}{mc^{2}R_{0}^{2}}\right)^{2} V^{2}}} = \frac{1}{\sqrt{1 - \gamma_{0} V^{2}}},
\end{equation}
where the dimensionless parameter
\begin{eqnarray}
	\gamma_{0} &=& \left(\frac{eM}{mc^{2}R_{0}^{2}}\right)^{2}=
	2.196 \times 10^{5} \times \left(\frac{m}{m_{e}}\right)^{-2} \nonumber\\
	&& 
	\times \left(\frac{M}{8 \times 10^{25}\;\mathrm{G\,cm^{3}}}\right)^{2}
	\times \left(\frac{R_{0}}{10^{10}\;\mathrm{cm}}\right)^{-4},
\end{eqnarray}
with $m_{e}=9.109\times 10^{-28}\,{\rm g}$ the electron mass.  The requirement that $\gamma$ be real imposes the kinematic bound
\begin{equation}
	V^{2} < \frac{1}{\gamma_{0}},
\end{equation}
which limits the maximum speed a particle can attain while still being treated in the relativistic dipole field model.

Finally, it is often advantageous for numerical integration to work directly with the velocity components.  Differentiating the definitions $V_X = dX/d\tau$, $V_Y = dY/d\tau$, $V_Z = dZ/d\tau$ gives the first-order system
\begin{eqnarray}
	\frac{dV_X}{d\tau} &=& \frac{1}{\gamma R^{3}}\left[-V_Y + 3\frac{Z}{R^{2}}\bigl(Z V_Y - Y V_Z\bigr)\right],\label{eq:VX}\\
	\frac{dV_Y}{d\tau} &=& \frac{1}{\gamma R^{3}}\left[V_X - 3\frac{Z}{R^{2}}\bigl(Z V_X - X V_Z\bigr)\right],\label{eq:VY}\\
	\frac{dV_Z}{d\tau} &=& 3\frac{1}{\gamma}\frac{Z}{R^{5}}\bigl(Y V_X - X V_Y\bigr).\label{eq:VZ}
\end{eqnarray}
These equations form a closed set of six first-order ordinary differential equations for $\{X,Y,Z,V_X,V_Y,V_Z\}$ that fully describe the relativistic motion of a charged particle in a dipole magnetic field.

\subsubsection{Radiative power loss}

The relativistic generalization of Larmor's formula for the power radiated by a transversely accelerated particle is
\begin{equation}
	P = \frac{2}{3}\frac{e^{2}}{c^{3}}\,\gamma^{4}\left(\frac{d\vec{v}}{dt}\right)^{2}.
\end{equation}
In the dimensionless variables introduced above, the emitted power can be expressed as
\begin{eqnarray}
	P &=& \frac{2}{3}\frac{e^{4}M^{2}}{m^{2}c^{5}R_{0}^{6}}\,\gamma^{4}
	\left[ \left(\frac{d^{2}X}{d\tau^{2}}\right)^{2} + \left(\frac{d^{2}Y}{d\tau^{2}}\right)^{2} + \left(\frac{d^{2}Z}{d\tau^{2}}\right)^{2} \right] \nonumber\\
	&=& P_{0}\,\gamma^{4}\,\tilde{P},
\end{eqnarray}
where
\begin{equation}
	P_{0} = \frac{2}{3}\frac{e^{4}M^{2}}{m^{2}c^{5}R_{0}^{6}},\qquad
	\tilde{P} = \left(\frac{d^{2}X}{d\tau^{2}}\right)^{2} + \left(\frac{d^{2}Y}{d\tau^{2}}\right)^{2} + \left(\frac{d^{2}Z}{d\tau^{2}}\right)^{2}.
\end{equation}

To obtain the power spectrum of the emitted radiation we compute the Discrete Fourier Transform (DFT) of the emitted power time series $P_k$ ($k=1,\dots,n$),
\begin{equation}
	P(\nu_z) = \frac{1}{\sqrt{n}}\sum_{k=1}^{n} P_k\, e^{2\pi i (k-1)(z-1)/n},
\end{equation}
and define the spectral power as $P(\nu) = |P(\nu_z)|^{2}$.

\section{Relativistic trajectories and radiation emission in CRSP: Exact solutions}\label{sect2}

In this section we analyze the trajectories and the emitted radiation power in the Classical Relativistic St\"{o}rmer Problem (CRSP), i.e., we study the solutions of the system (\ref{F1})--(\ref{F3}) and their properties. While approximate perturbative solutions can be constructed, we shall restrict ourselves to a numerical approach for the full three-dimensional dynamics. Nevertheless, we begin by presenting an exact solution of the equations of motion in the equatorial plane.

\subsection{Exact solutions of the relativistic St\"{o}rmer problem}

Restricting the motion to the equatorial plane $Z=0$ reduces the equations of motion (\ref{F1})--(\ref{F3}) to
\begin{equation}
	\frac{d^{2}X}{d\tau^{2}} = -\frac{1}{\gamma}\frac{1}{R^{3}}\frac{dY}{d\tau},
	\qquad
	\frac{d^{2}Y}{d\tau^{2}} = \frac{1}{\gamma}\frac{1}{R^{3}}\frac{dX}{d\tau},
	\label{F4}
\end{equation}
which admit the first integral
\begin{equation}
	V^{2} = \left(\frac{dX}{d\tau}\right)^{2} + \left(\frac{dY}{d\tau}\right)^{2} = \mathrm{constant}.
\end{equation}
For this planar motion the dimensionless electromagnetic power radiated by the particle simplifies to
\begin{equation}
	\tilde{P} = \frac{V^{2}}{\gamma^{2} R^{6}}.
\end{equation}

%\subsubsection{The simple harmonic oscillator solution}

A particularly transparent solution of (\ref{F4}) is obtained by assuming uniform circular motion,
\begin{equation}
	X(\tau) = -R_{0}\sin(\omega\tau+\phi), \qquad
	Y(\tau) =  R_{0}\cos(\omega\tau+\phi),
\end{equation}
where
\begin{equation}
	R_{0} = \sqrt{X_{0}^{2}+Y_{0}^{2}}, \qquad
	\tan\phi = -\frac{X_{0}}{Y_{0}},
\end{equation}
and $(X_{0},Y_{0})$ are the initial coordinates. The angular frequency $\omega$ is tied to the orbital radius via $\omega = 1/\gamma R_{0}^{3}$. Moreover, one finds the relations $\omega^{2} R_{0}^{2}=V^{2}$ and $V = 1/(\gamma R_{0}^{2}) = 1/(\gamma (X_{0}^{2}+Y_{0}^{2}))$. This solution describes simple harmonic oscillations along the $X$ and $Y$ axes, i.e., a perfectly circular trajectory of constant dimensionless radius $R_{0}$.

The emitted electromagnetic power for this orbit is
\begin{equation}
	P = P_{0}\gamma^{4} R_{0}^{2}\omega^{4} = P_{0}\gamma^{4} V^{2}\omega^{2}.
\end{equation}
The trajectory of the particle is strictly circular, with constant dimensionless radius $R_0$. 
The dimensionless period of the motion is $\tilde{T} = 2\pi\gamma R_{0}^{3}$, which is proportional to the Lorentz factor $\gamma$. The physical period, expressed in coordinate time, reads
\begin{equation}
	T = \frac{2\pi mc R_{0}^{3}}{eM}\,\gamma R_{0}^{3} = \frac{2\pi mc\gamma R_{0}^{6}}{eM},
\end{equation}
and is thus inversely proportional to the magnetic moment $M$ of the central object.

\subsection{An exact parametric solution of the CRSP}

The simple harmonic oscillator solution found above is a special case of a wider class of planar orbits in the equatorial plane. To uncover this broader family we look for solutions of (\ref{F4}) in the polar form
\begin{equation}
	X(\tau) = R(\tau)\sin\omega(\tau), \qquad
	Y(\tau) = R(\tau)\cos\omega(\tau),
\end{equation}
so that $R(\tau) = \sqrt{X^{2}+Y^{2}}$ and $\omega(\tau) = \arctan(X/Y)$. This ansatz allows the orbital radius and angular coordinate to become time-dependent, while still respecting the planar nature of the motion.

Substituting these expressions into the equations of motion (\ref{F4}) yields a coupled system for the dynamical radius $R(\tau)$ and the angular function $\omega(\tau)$,
\begin{equation}
	R''(\tau) - \omega'^{2}(\tau) R(\tau) - \frac{\omega'(\tau)}{\gamma R^{2}(\tau)} = 0, \label{42}
\end{equation}
\begin{equation}
	R(\tau)\,\omega''(\tau) + 2R'(\tau)\,\omega'(\tau) + \frac{R'(\tau)}{\gamma R^{3}(\tau)} = 0, \label{43}
\end{equation}
where a prime denotes differentiation with respect to the dimensionless time $\tau$. Equation (\ref{43}) can be rearranged as
\begin{equation}
	\frac{1}{R(\tau)}\frac{d}{d\tau}\!\Bigl[R^{2}(\tau)\,\omega'(\tau)\Bigr] + \frac{R'(\tau)}{\gamma R^{3}(\tau)} = 0,
\end{equation}
which is immediately integrable and gives
\begin{equation}
	\omega'(\tau) = \frac{1}{R^{2}(\tau)}\Bigl[C + \frac{1}{\gamma R(\tau)}\Bigr]. \label{45}
\end{equation}
The integration constant $C$ can be interpreted physically: when $C$ vanishes, the angular velocity reduces to $\omega' = 1/(\gamma R^{3})$, which is precisely the frequency of the simple circular orbit; a nonzero $C$ therefore encodes a deviation from that purely harmonic motion. The system (\ref{F4}) also admits a first integral corresponding to the conservation of the dimensionless speed,
\begin{equation}
	R'^{2}(\tau) + R^{2}(\tau)\,\omega'^{2}(\tau) = V^{2} = \mathrm{constant}. \label{44}
\end{equation}

Inserting $\omega'(\tau)$ from (\ref{45}) into the energy integral (\ref{44}) decouples the radial motion:
\begin{equation}
	R'^{2}(\tau) + \frac{1}{R^{2}(\tau)}\Bigl[C + \frac{1}{\gamma R(\tau)}\Bigr]^{2} = V^{2}. \label{46}
\end{equation}
This is a first-order equation for $R(\tau)$ whose right-hand side can be interpreted as an effective radial potential
\begin{equation}
	V_{\rm eff}(R) = \frac{1}{R^{2}}\Bigl(C + \frac{1}{\gamma R}\Bigr)^{2},
\end{equation}
so that the motion in $R$ is that of a particle of constant energy $V^{2}$ moving in this potential. Physically admissible radii are those for which $V^{2} \ge V_{\rm eff}(R)$; the turning points occur where equality holds, and the available radial range can be read directly from the shape of $V_{\rm eff}$.

The exact solution for $R(\tau)$ is obtained implicitly from (\ref{46}):
\begin{equation}
	\tau(R) - \tau_{0} = \int_{R_{0}}^{R} \frac{d\xi}{\sqrt{\,V^{2} - \frac{1}{\xi^{2}}\bigl(C + \frac{1}{\gamma\xi}\bigr)^{2}\,}}, \label{47}
\end{equation}
where $R_{0}=R(\tau_{0})$. This integral can be evaluated in closed form in terms of elliptic functions. Introducing the auxiliary variable $\theta$ through
\begin{equation}
	\frac{1}{\xi}\Bigl(C + \frac{1}{\gamma\xi}\Bigr) = V\sin\theta,
\end{equation}
one finds
\begin{equation}
	\xi = \frac{C}{2V}\Bigl[ \bigl(1 + \sqrt{1 + \tfrac{4V}{\gamma C^{2}}\sin\theta}\,\bigr) \csc\theta \Bigr],
\end{equation}
which transforms (\ref{47}) into
\begin{eqnarray}
	&&\tau(R) - \tau_{0} = -\frac{C}{2V^{2}} \times
		\nonumber \\
	&&\int_{\theta_{0}}^{\theta}
	\frac{\csc^{2}\theta\;\Bigl(1 + \sqrt{1 + \tfrac{4V}{\gamma C^{2}}\sin\theta} + \tfrac{2V}{\gamma C^{2}}\sin\theta\Bigr)\,d\theta}
	{\sqrt{1 + \tfrac{4V}{\gamma C^{2}}\sin\theta}}.
\end{eqnarray}
Carrying out the integration gives the explicit parametric relation
\begin{equation}
	\tau(R) - \tau_{0} = \bigl.F(\theta)\bigr|_{\theta_{0}(R_{0})}^{\theta(R)},
\end{equation}
where
\begin{eqnarray}
	F(\theta) &=& \frac{C}{2V^{2}\sqrt{1 + \tfrac{4V}{\gamma C^{2}}\sin\theta}} \nonumber\\
	&&\times \Bigg[ \cot\theta\Bigl( \sqrt{1 + \tfrac{4V}{\gamma C^{2}}\sin\theta} + \gamma C\Bigr) \nonumber\\
	&&\quad + \sqrt{\frac{1 + \tfrac{4V}{\gamma C^{2}}\sin\theta}{1 + \tfrac{4V}{\gamma C^{2}}}}\,
	\mathrm{F}\!\left( \frac{\pi - 2\theta}{4} \Bigg| \frac{8V}{\gamma C^{2} + 4V} \right) \nonumber\\
	&&\quad - \sqrt{\bigl(1 + \tfrac{4V}{\gamma C^{2}}\bigr)\bigl(1 + \tfrac{4V}{\gamma C^{2}}\sin\theta\bigr)}\times
		\nonumber \\
	&&\times \mathrm{E}\!\left( \frac{\pi - 2\theta}{4} \Bigg| \frac{8V}{\gamma C^{2} + 4V} \right)  + 4V\cos\theta \Bigg],
\end{eqnarray}
and $\mathrm{F}(\phi|m)$, $\mathrm{E}(\phi|m)$ are the incomplete elliptic integrals of the first and second kind. The limits are
\begin{eqnarray}
	\theta_{0}(R_{0}) &=& \arcsin\!\left[ \frac{1}{V R_{0}}\Bigl(C + \frac{1}{\gamma R_{0}}\Bigr) \right], \\
	\theta(R)         &=& \arcsin\!\left[ \frac{1}{V R}\Bigl(C + \frac{1}{\gamma R}\Bigr) \right].
\end{eqnarray}

In a similar manner, the angular function $\omega(R)$ can be obtained by combining
\begin{equation}
	\frac{d\omega}{d\tau} = \frac{d\omega}{dR}\frac{dR}{d\tau}
\end{equation}
with (\ref{46}), yielding
\begin{equation}
	\omega(R) - \omega_{0} = \int_{R_{0}}^{R} \frac{\bigl(C + \frac{1}{\gamma\xi}\bigr)\,d\xi}
	{\xi^{2}\sqrt{\,V^{2} - \frac{1}{\xi^{2}}\bigl(C + \frac{1}{\gamma\xi}\bigr)^{2}\,}}, \label{48}
\end{equation}
where $\omega_{0}$ is the initial angular coordinate. Changing variable to $\sigma = 1/\xi$ gives
\begin{equation}
	\omega(R) - \omega_{0} = -\int_{1/R_{0}}^{1/R} \frac{(C + \sigma/\gamma)\,d\sigma}
	{\sqrt{\,V^{2} - \sigma^{2}\bigl(C + \sigma/\gamma\bigr)^{2}\,}}.
\end{equation}
Introducing $\psi$ via $\sigma\bigl(C + \sigma/\gamma\bigr) = V\sin\psi$, so that
\begin{equation}
	\sigma = \frac{1}{2}\,\gamma C\Bigl( \sqrt{1 + \tfrac{4V}{\gamma C^{2}}\sin\psi} - 1\Bigr),
\end{equation}
the integral becomes
\begin{equation}
	\omega(R) - \omega_{0} = -\frac{1}{2}\int_{\psi_{0}}^{\psi}
	\Bigl( 1 + \frac{1}{\sqrt{\tfrac{4V}{\gamma C^{2}}\sin\psi}} \Bigr)\,d\psi,
\end{equation}
which is again expressible in terms of elliptic integrals:
\begin{equation}
	\omega(R) - \omega_{0} = \bigl.G(\psi)\bigr|_{\psi_{0}(R_{0})}^{\psi(R)},
\end{equation}
with
\begin{equation}
	G(\psi) = \frac{ \sqrt{1 + \tfrac{4V}{\gamma C^{2}}\sin\psi}\;
		\mathrm{F}\!\left( \frac{\pi - 2\psi}{4} \Bigg| \frac{8V}{\gamma C^{2} + 4V} \right) }
	{ \sqrt{1 + \tfrac{4V}{\gamma C^{2}}}\;\sqrt{1 + \tfrac{4V}{\gamma C^{2}}\sin\psi} }
	- \frac{\psi}{2},
\end{equation}
and
\begin{eqnarray}
	\psi_{0}(R_{0}) &=& \arcsin\!\left[ \frac{1}{V R_{0}}\Bigl(C + \frac{1}{\gamma R_{0}}\Bigr) \right],
		 \\
	\psi(R) &=& \arcsin\!\left[ \frac{1}{V R}\Bigl(C + \frac{1}{\gamma R}\Bigr) \right].
\end{eqnarray}

Equations~(\ref{47}) and~(\ref{48}) together constitute the exact solution of the planar relativistic St\"{o}rmer problem, with the instantaneous radius $R$ playing the role of a parameter. The trajectory is therefore given in parametric form: as $R$ varies between its turning points, $\tau(R)$ and $\omega(R)$ trace out the orbit. The solution is physical for those values of $R$ for which
\begin{equation}
	V > \left|\,\frac{1}{\xi}\Bigl(C + \frac{1}{\gamma\xi}\Bigr)\right|_{\xi=R_{0}}^{\xi=R},
\end{equation}
guaranteeing that the radial momentum remains real.  

The solution depends on two parameters $(V,C)$, and on the initial values $R_0$ and $\omega _0$ of $R$ and $\omega$. The initial values of the solution are fixed by the initial values of $X$ and $Y$, $X(0)=X_0$ and $Y(0)=Y_0$ by the relations $R_0=R(0)=\sqrt{X_0^2+Y_0^2}$, and $\omega (0)=\arctan\left(X_0/Y_0\right)$, respectively.

\subsubsection{The extreme relativistic limit}

\begin{figure*}[!htbp]
	\includegraphics[scale=0.5]{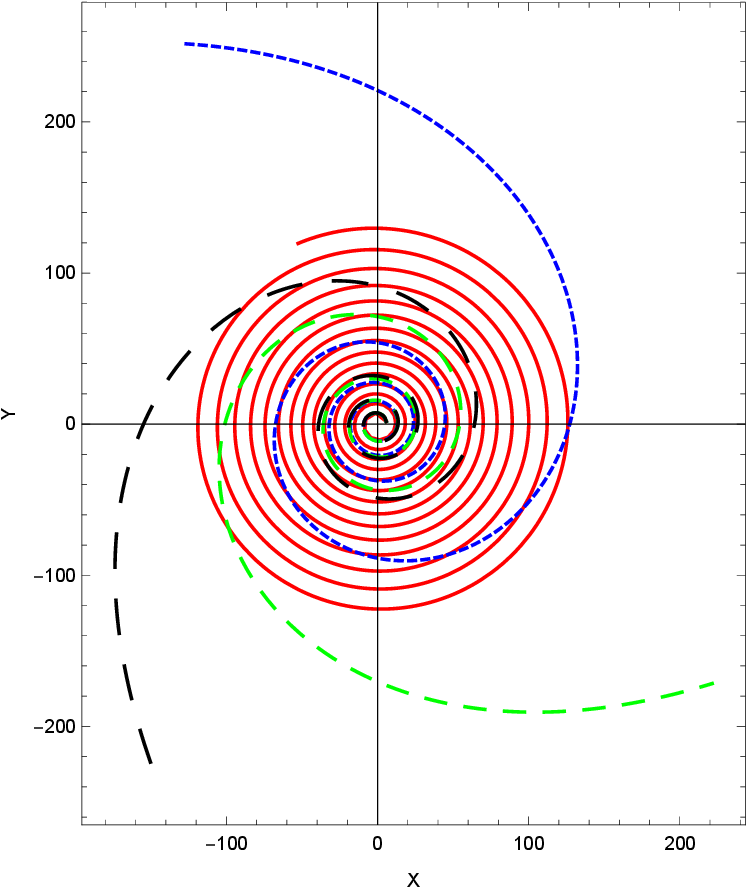}\\
	\includegraphics[scale=0.58]{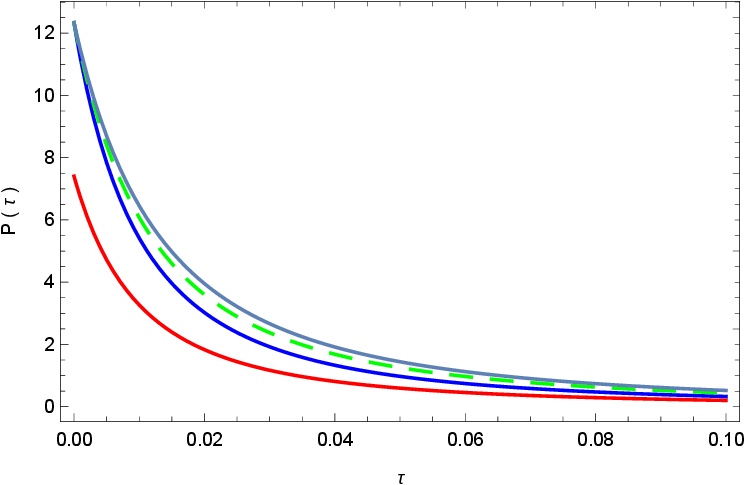}\qquad \qquad
	\includegraphics[scale=0.58]{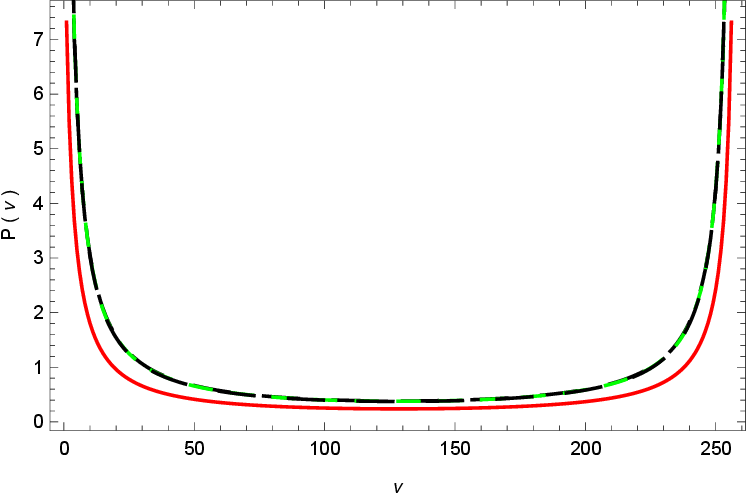}
	\caption{Trajectory, radiation power, and radiation spectrum obtained from the exact solution of the two-dimensional Classical Relativistic St\"{o}rmer Problem (CRSP) for $\vec{R}_0=(1.7,0.9)$, $R(0)=R_0=1.92$, $\omega (0)=\omega _0=1.083$, $\gamma =2$,  and different values of $C$ and $V$: $C=-120,V=0.10$ (solid red curve), $C=-120,V=0.50$ (dotted blue curve), $C=-100, V=0.50$ (dashed green curve), and $C=-90,V=0.50$ (black large dashed curve), respectively. \label{Fig0}}
\end{figure*} 

When $\gamma \gg 1$, the term $1/(\gamma \xi)$ in the effective potential becomes negligible compared to the constant $C$ (assuming $C\neq 0$). In this ultra-relativistic regime the integral (\ref{47}) for the radial motion simplifies to
\begin{equation}
	\tau(R)-\tau_0 = \int_{R_0}^{R} \frac{d\xi}{\sqrt{V^2 - \frac{C^2}{\xi^2}}},
\end{equation}
which can be evaluated in closed form. The integration yields
\begin{equation}
	R(\tau) = \pm\sqrt{ 2 R_0 (\tau-\tau_0)\sqrt{V^2 - \frac{C^2}{R_0^2}} + R_0^2 + V^2 (\tau-\tau_0)^2 }.
	\label{eq:rad_extreme}
\end{equation}

\begin{figure*}[!htbp]
	\includegraphics[scale=0.5]{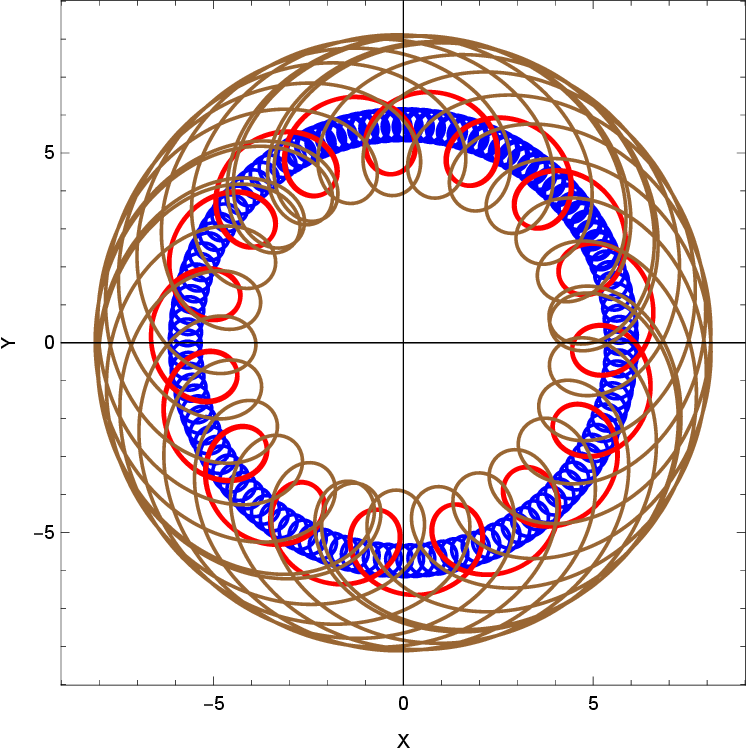}\\
	\includegraphics[scale=0.58]{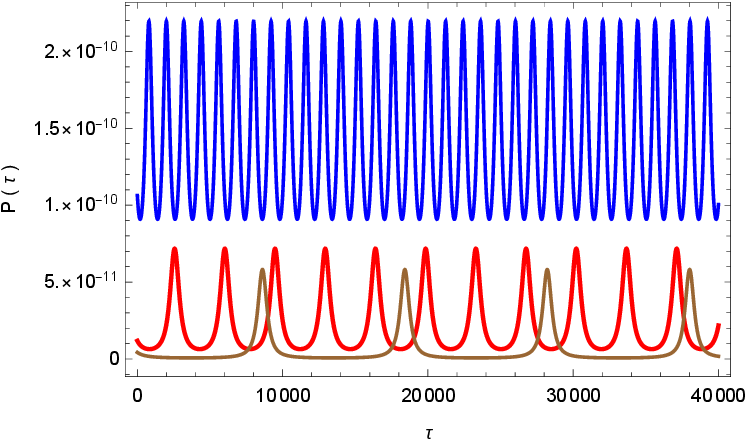}\qquad \qquad
	\includegraphics[scale=0.58]{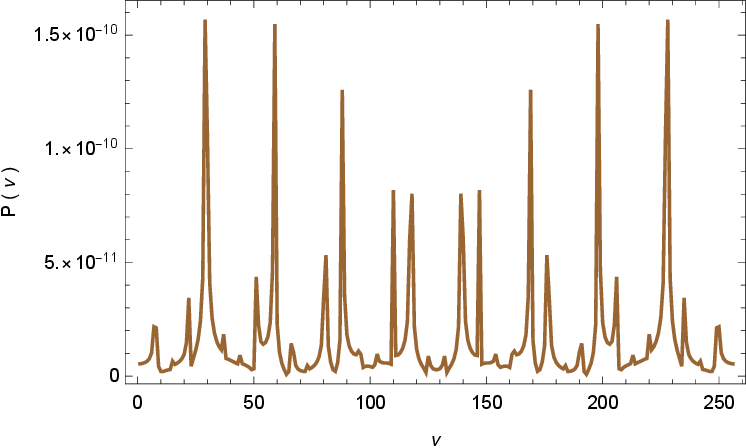}
	\caption{Trajectories, radiation powers, and radiation spectrum of the Classical Relativistic St\"{o}rmer Problem (CRSP) for $\vec{R}_0=(5.7,1.9,0)$, $R(0)=6.608$, $\vec{V}_0=\left(0.001,0.002,0\right)$, $V(0)=0.0022$, and for different values of $\gamma$: $\gamma =1$ (solid blue curve), $\gamma =3$ (solid red curve), and $\gamma =5$  (solid brown curve), respectively. The radiation spectrum is presented only for the case $\gamma =5$. \label{Fig0a}}
\end{figure*} 

From this expression one sees that at late times the radial coordinate grows linearly, $R(\tau) \sim V(\tau-\tau_0)$, indicating that the particle ultimately escapes to infinity; no bounded motion is possible because the effective potential is purely repulsive $C^2/R^2$ and lacks the barrier that existed for $\gamma \lesssim 1$.

Similarly, the angular variable obeys
\begin{equation}
	\omega(R) - \omega_0 = C \int_{R_0}^{R} \frac{d\xi}{\xi^2 \sqrt{V^2 - \frac{C^2}{\xi^2}}},
\end{equation}
leading to
\begin{equation}
	\omega(R) = \omega_0 + \arctan\frac{C}{\sqrt{V^2 R_0^2 - C^2}} - \arctan\frac{C}{\sqrt{V^2 R^2 - C^2}}.
	\label{eq:ang_extreme}
\end{equation}
Equation (\ref{eq:ang_extreme}) shows that the angular coordinate approaches a finite limit as $R\to\infty$, so the trajectory approaches a straight line at a fixed angle, consistent with an unbound particle moving essentially along a radial path in the equatorial plane. Thus, in the extreme relativistic limit the planar motion becomes unbound, in sharp contrast to the rich bound orbits available when $\gamma$ is moderate.

\begin{figure*}[t]
	\includegraphics[scale=0.54]{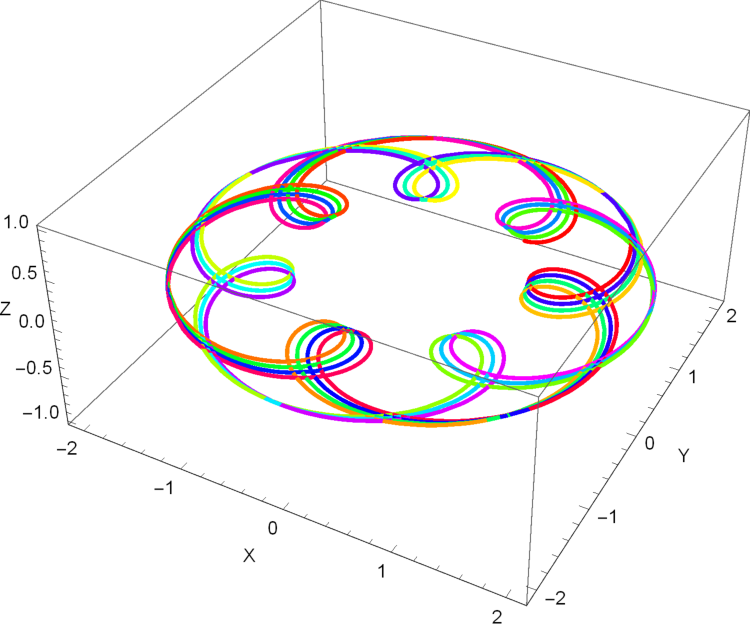}\\
	\includegraphics[scale=0.58]{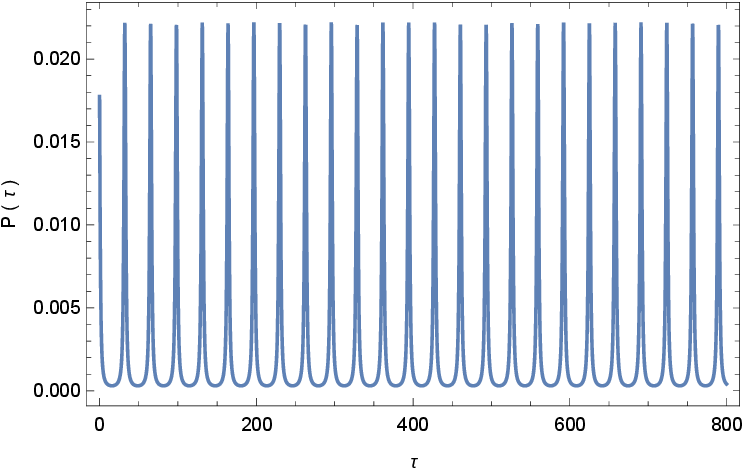}\qquad \qquad
	\includegraphics[scale=0.58]{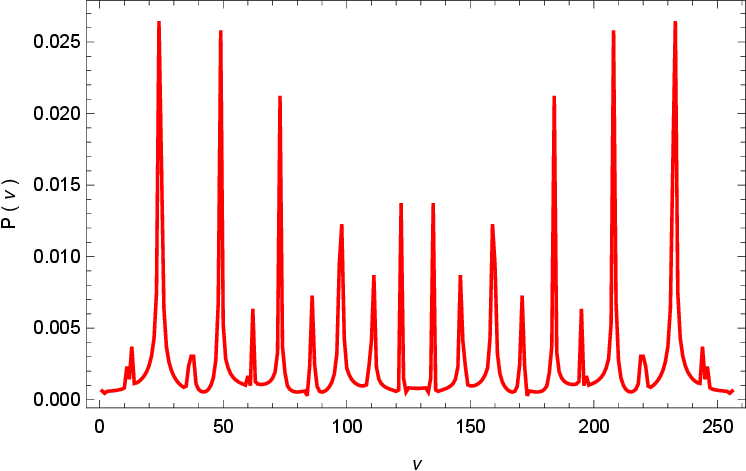}
	\caption{Trajectory, radiation power, and radiation spectrum for the Classical St\"{o}rmer Problem (CSP) for $\vec{R}_0=(0.7,0.8,0)$, $\left|\vec{R}_0\right| =1.063$,  $\vec{V}_0=(0.16,0,0)$, and $\gamma =1$. \label{Fig1}}
\end{figure*}

\begin{figure*}[!htbp]
	\includegraphics[scale=0.54]{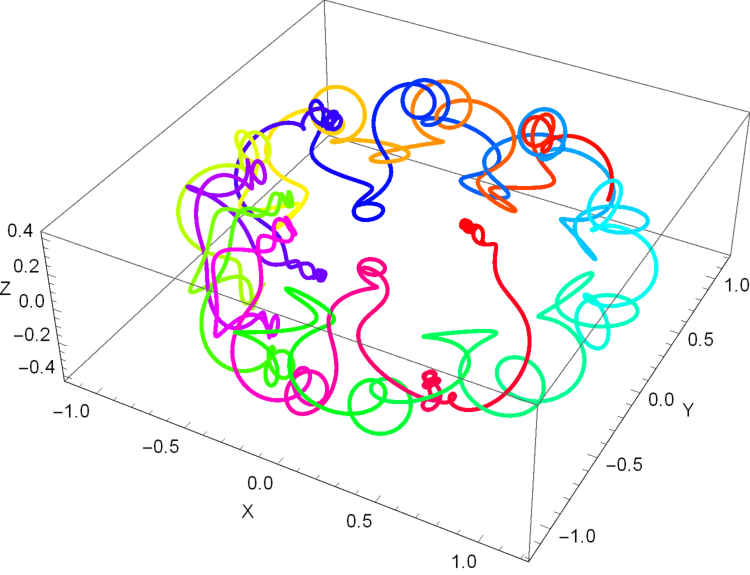}\\
	\includegraphics[scale=0.58]{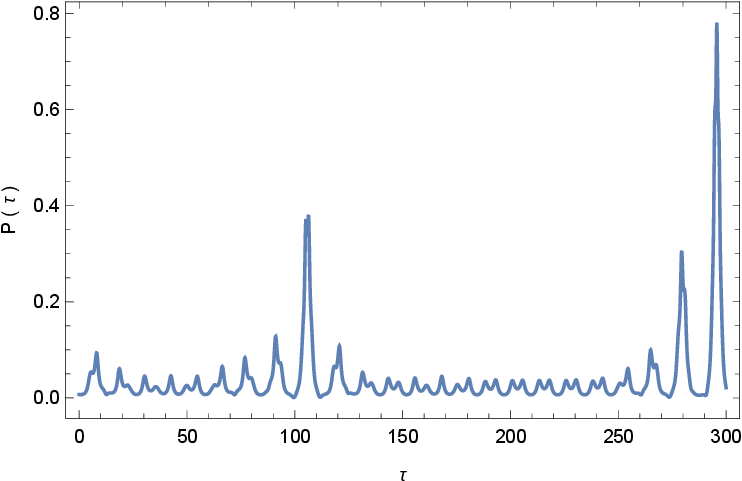}\qquad \qquad
	\includegraphics[scale=0.58]{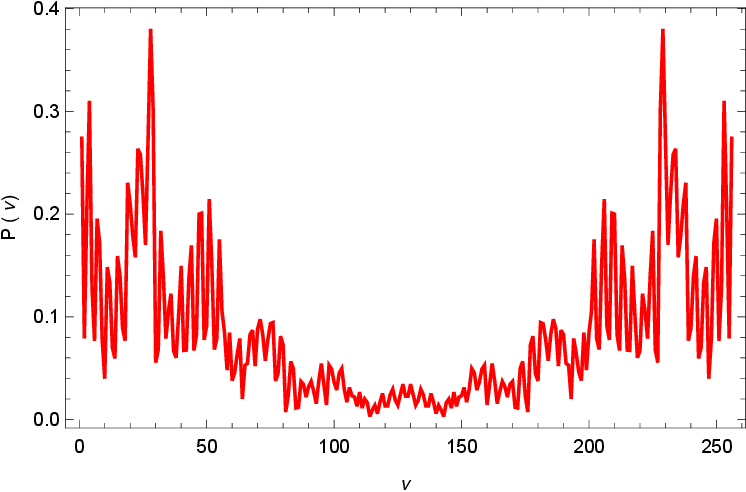}
	\caption{Trajectory, radiation power, and radiation spectrum for the Classical St\"{o}rmer Problem (CSP) for $\vec{R}_0=(0.7,0.8,0)$, $\left|\vec{R}_0\right| =1.063$, $\vec{V}_0=(0.01,0,10,0.10)$ and $\gamma =1$. \label{Fig2}}
\end{figure*}

\begin{figure*}[!htbp]
	\includegraphics[scale=0.54]{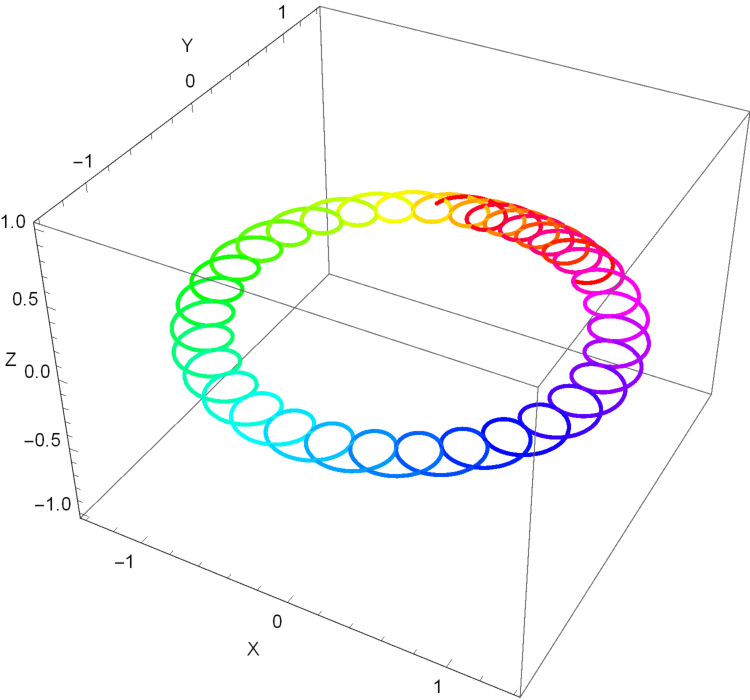}\\
	\includegraphics[scale=0.58]{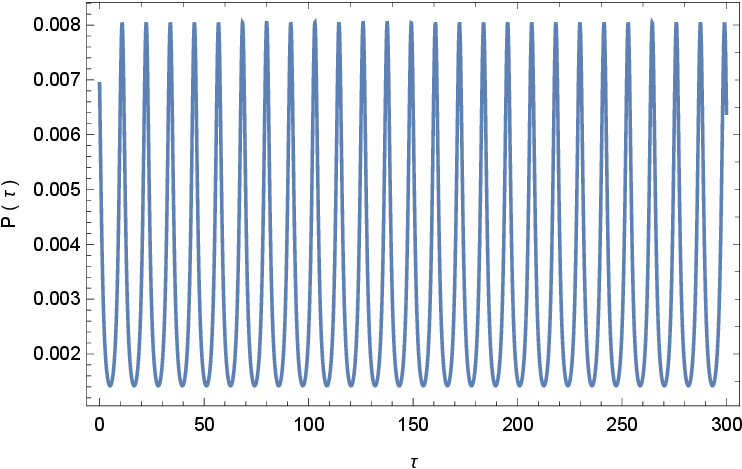}\qquad \qquad
	\includegraphics[scale=0.58]{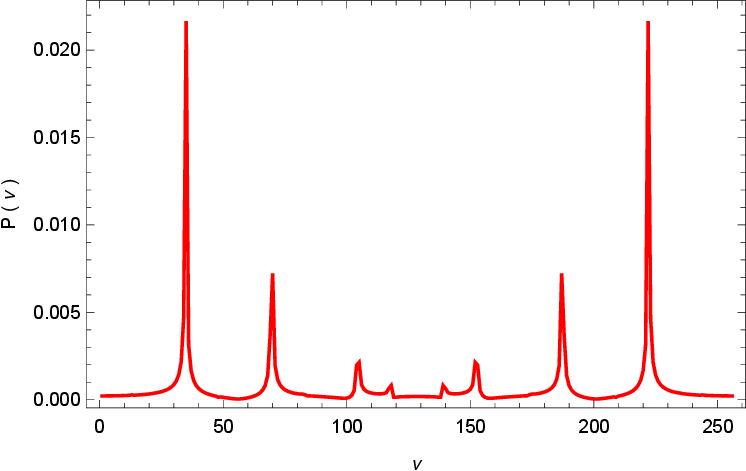}
	\caption{Trajectory, radiation power, and radiation spectrum for the Classical St\"{o}rmer Problem (CSP) for $\vec{R}_0=(0.7,0.8,0)$, $\left|\vec{R}_0\right| =1.063$,  $\vec{V}_0=(0.10,0,0)$, and $\gamma =1$. \label{Fig3}}
\end{figure*}

\subsubsection{The case $C \ll 1/(\gamma \xi)$}

If the integration constant $C$ is chosen to be negligibly small compared to the term $1/(\gamma \xi)$, the dynamics simplify considerably. Physically, this limit describes particles for which the ``centrifugal'' contribution associated with $C$ is dominated by the $\gamma$-dependent part of the effective potential. In this regime the radial motion reduces to
\begin{equation}\label{47a}
	\tau(R)-\tau_0 = \int_{R_0}^{R} \frac{d\xi}{\sqrt{V^{2} - \dfrac{1}{\gamma^{2}\xi^{4}}}}.
\end{equation}
Introducing the new variable $\sigma = 1/\xi$, the integral becomes
\begin{equation}
	\tau(R)-\tau(R_0) = -\int_{1/R_0}^{1/R} \frac{d\sigma}{\sigma^{2}\sqrt{V^{2} - \sigma^{4}/\gamma^{2}}},
\end{equation}
and can be evaluated in closed form in terms of the Gauss hypergeometric function. The result is
\begin{eqnarray}
	&&\tau (R)-\tau _0  =\nonumber\\
	&&\frac{\sqrt{V^{2}-\frac{1}{R^{4}\gamma
				^{2}}}\left( \,_{2}F_{1}\left( 1,\frac{5}{4};\frac{7}{4};\frac{1}{%
			R^{4}V^{2}\gamma ^{2}}\right) +3R^{4}\gamma ^{2}V^{2}\right) }{3R^{3}\gamma
		^{2}V^{4}} \nonumber\\
	&&-\frac{\sqrt{V^{2}-\frac{1}{R_{0}^{4}\gamma ^{2}}}\left( \,_{2}F_{1}\left(
		1,\frac{5}{4};\frac{7}{4};\frac{1}{R_{0}^{4}V^{2}\gamma ^{2}}\right)
		+3R_{0}^{4}\gamma ^{2}V^{2}\right) }{3R_{0}^{3}\gamma ^{2}V^{4}},\nonumber\\
\end{eqnarray}
where ${}_{2}F_{1}(a,b;c;z) = \sum_{k=0}^{\infty} \frac{(a)_{k} (b)_{k}}{(c)_{k}} \frac{z^{k}}{k!}$ is the standard hypergeometric function. The angular variable is obtained from
\begin{equation}
	\omega(R)-\omega_0 = \int_{R_0}^{R} \frac{d\xi}{\gamma\,\xi^{3}\sqrt{V^{2} - \dfrac{1}{\gamma^{2}\xi^{4}}}},
\end{equation}
which yields the compact expression
\begin{eqnarray}
	\omega(R)-\omega_0 &=& \frac{1}{2}\tan^{-1}\!\left( \gamma R^{2}\sqrt{V^{2} - \frac{1}{\gamma^{2}R^{4}}}\right)
		\nonumber \\
	&&- \frac{1}{2}\tan^{-1}\!\left( \gamma R_{0}^{2}\sqrt{V^{2} - \frac{1}{\gamma^{2}R_{0}^{4}}}\right).
\end{eqnarray}

In this regime the effective potential $V_{\rm eff}\approx 1/(\gamma^{2}R^{4})$ is purely repulsive. Consequently the particle always follows an unbound trajectory: $R(\tau)$ grows without bound, and as $R\to\infty$ the arctangent in $\omega(R)$ approaches a constant, so the orbit rapidly becomes radially outward. This analytic solution therefore provides a faithful description of particles injected into the equatorial plane with negligible $C$ and sufficient energy to escape the trapping barriers that exist for larger $C$.

\subsubsection{Numerical analysis}

The radial motion described by Eq.~(\ref{46}) can be interpreted as the one-dimensional motion of a particle of unit mass in the effective potential
\begin{equation}
	V_{\rm eff}(R) = \frac{1}{2R^{2}}\Bigl[ C + \frac{1}{\gamma R}\Bigr]^{2},
	\label{eq:Veff}
\end{equation}
with conserved energy $\epsilon = V^{2}/2$, so that
\begin{equation}
	\frac{1}{2}R'^{2} + V_{\rm eff}(R) = \epsilon.
\end{equation}
The shape of $V_{\rm eff}$ controls whether the particle is trapped at finite distance or escapes to infinity.  The turning radii, where $R'=0$, satisfy $V_{\rm eff}(R) = \epsilon$, i.e.
\begin{equation}
	\frac{1}{2R^{2}}\Bigl( C + \frac{1}{\gamma R}\Bigr)^{2} = \frac{V^{2}}{2},
\end{equation}
which yields four formal solutions,
\begin{eqnarray}
	R_{1,2} &=& \frac{C}{2V}\Bigl( 1 \pm \sqrt{1 + \frac{4V}{\gamma C^{2}}}\Bigr), \\
	R_{3,4} &=& \frac{C}{2V}\Bigl( -1 \pm \sqrt{1 - \frac{4V}{\gamma C^{2}}}\Bigr).
	\label{eq:turning}
\end{eqnarray}
Only real and positive values correspond to physical turning points.  When $C<0$ and $V < \gamma C^{2}/4$, the potential develops a barrier, leading to two turning points $R_{2}$ and $R_{4}$ (with $R_{4}<R_{2}$) that bound a finite interval; when $V$ exceeds the barrier height, only one turning point exists on the low-$R$ side and the particle eventually escapes.

Extrema of the effective potential correspond to circular orbits in the equatorial plane and are obtained from $dV_{\rm eff}/dR=0$:
\begin{equation}
	\frac{dV_{\rm eff}}{dR} = -\frac{(\gamma C R + 1)(\gamma C R + 2)}{\gamma^{2} R^{5}} = 0,
\end{equation}
giving the two equilibrium radii
\begin{equation}
	R_{\rm eq}^{(1)} = -\frac{2}{\gamma C}, \qquad
	R_{\rm eq}^{(2)} = -\frac{1}{\gamma C}.
\end{equation}
The stability of these equilibria is determined by the second derivative,
\begin{equation}
	\frac{d^{2}V_{\rm eff}}{dR^{2}} = \frac{3\gamma C R(\gamma C R + 4) + 10}{\gamma^{2} R^{6}},
\end{equation}
evaluated at each point:
\begin{equation}
	\left.\frac{d^{2}V_{\rm eff}}{dR^{2}}\right|_{R=R_{\rm eq}^{(1)}} = -\frac{1}{32}\,\gamma^{4} C^{6}, \qquad
	\left.\frac{d^{2}V_{\rm eff}}{dR^{2}}\right|_{R=R_{\rm eq}^{(2)}} = \gamma^{4} C^{6}.
\end{equation}
Hence $R_{\rm eq}^{(1)}$ is a local maximum (unstable circular orbit) and $R_{\rm eq}^{(2)}$ is a local minimum (stable circular orbit).  The unstable orbit sits on top of the potential barrier and separates bound motion within the well from unbound trajectories that escape to large distances.

To illustrate the behaviour of exact solutions, we present in Fig.~\ref{Fig0} a representative set of trajectories, the emitted electromagnetic power, and the corresponding Fourier power spectrum, obtained for a fixed initial position $(X_{0},Y_{0})$ and initial angle $\omega_{0}$, with Lorentz factor $\gamma = 2$.  The exact solution typically describes spiralling trajectories in which the radial distance $R(\tau)$ increases with time; for the chosen parameters the particle escapes the central magnetic field and moves out to very large distances.  The radiated power decreases rapidly as $R$ grows, consistent with the $R^{-6}$ scaling of the dimensionless power $\tilde{P}$.  The power spectrum adopts a very simple form, dominated by two prominent peaks that reflect the fundamental frequencies of the quasi-periodic motion before escape.

\section{Relativistic trajectories and radiation emission in CRSP-numerical results}\label{sect3}

In the general case, the study of the St\"{o}rmer problem can be done by using numerical methods only. The nature of the solution, and correspondingly the nature of the motion of the particles in the dipole magnetic field essentially depends on the set of the initial conditions $\left(\vec{R}_0,\vec{V}_0\right)$ of the particle. In the following we will consider the numerical solutions of both the CSP and of the CRSP corresponding to different initial configurations of the motion.

\subsection{Two-dimensional planar motions} 

We begin our numerical investigations on the motion of charged particles in the dipole magnetic field with the planar case, corresponding to $Z=0$, and with the motion of the particle confined to the two-dimensional geometry of a plane. The particle orbits, the emitted power, and one example of radiation spectrum are represented in Fig.~\ref{Fig0a}, for three different values of the Lorentz factor $\gamma$.  The case $\gamma =1$ corresponds to the non-relativistic motion of the particle, represented by the blue curve in the Figure. The motion is periodic, regular, and stable, and the orbits are confined in a small region of the magnetic field.  With the increase of $\gamma$ to the value $\gamma =3$, and with the same initial conditions, the shape of the orbit changes, with the particle covering a larger area in its motion. However, the motion is still characterized by a regular and periodic pattern. For $\gamma =5$, the nature of the trajectory changes significantly, it increases in complexity, and the motion of the particle extends to a still larger area. For values of $\gamma $ larger than 5, for the given initial conditions, the particle escapes to infinity. 

The emitted electromagnetic power, represented in the middle panel of Fig.~\ref{Fig0a}, shows a periodic and regular emission  pattern, with the intensity of the radiation decreasing with increasing $\gamma$. The highest radiation emission power corresponds to the Newtonian case, with $\gamma =1$.  There is no dissipation in the signal, and its amplitude is a constant, since the effects of the energy losses and of the radiation reaction are neglected in the equation of motion. The radiation  spectrum, presented only for the case $\gamma =5$, indicates the presence of sharp peaks in the spectrum, as well as a periodic structure in the spectral distribution. The emission spectrum represents a distinct observational signature of the particle motion in the dipole magnetic field, and it is highly dependent on the presence of the relativistic effects.

\subsection{Non-relativistic 3D motions}

As a next step we explore the non-relativistic limit of three-dimensional motion, obtained by setting $\gamma \to 1$. Several representative trajectories and the associated radiation patterns are shown, for different initial conditions, in Figs.~\ref{Fig1}--\ref{Fig3}. In all three figures the particle is initially placed in the equatorial plane, $Z_0 = 0$, while its initial velocity may have arbitrary out-of-plane components. In Figs.~\ref{Fig1} and \ref{Fig3} the initial velocity is directed purely along the $X$ axis; the resulting motion is three-dimensional but remains highly regular. The emitted electromagnetic power is strictly periodic with constant amplitude, whose value depends on the chosen initial conditions. Despite the regularity of the signal, the power spectrum for each case displays characteristic features that constitute distinct signatures of the radiation emission.

When all three components of the initial velocity are non-zero (Fig.~\ref{Fig2}) the dynamics become irregular. The particle trajectory departs from any simple periodicity and shows a complex pattern, suggestive of the onset of chaotic motion. The radiated power is no longer a clean periodic signal; instead it consists of bursts of emission superimposed on an approximately constant background. The corresponding power spectrum retains sharp peaks, but their arrangement reflects the underlying irregular motion. This contrast illustrates how sensitive the CRSP dynamics are to the initial velocity, even in the non-relativistic regime.

\subsection{Relativistic motions}

To integrate the equations of the relativistic St\"{o}rmer problem numerically one must specify the initial position $\vec{R}_0$, the initial velocity $\vec{V}_0$, and the Lorentz factor $\gamma$. The initial speed is tied to $\gamma$ by
\begin{equation}
	V_0 = \left(\frac{eM}{mc^{2}R_0^{2}}\right)^{-1}\frac{\sqrt{\gamma^{2}-1}}{\gamma},
\end{equation}
so that fixing $\gamma$ and $V_0$ is physically equivalent to assigning a value to the ratio $M/R_0^{2}$, i.e. to setting the magnetic structure of the dipole field. Thus the pair $\left(\gamma, V_0\right)$ fully determines the magnetic environment in which the particle moves.

\begin{figure*}[!htbp]
	\includegraphics[scale=0.54]{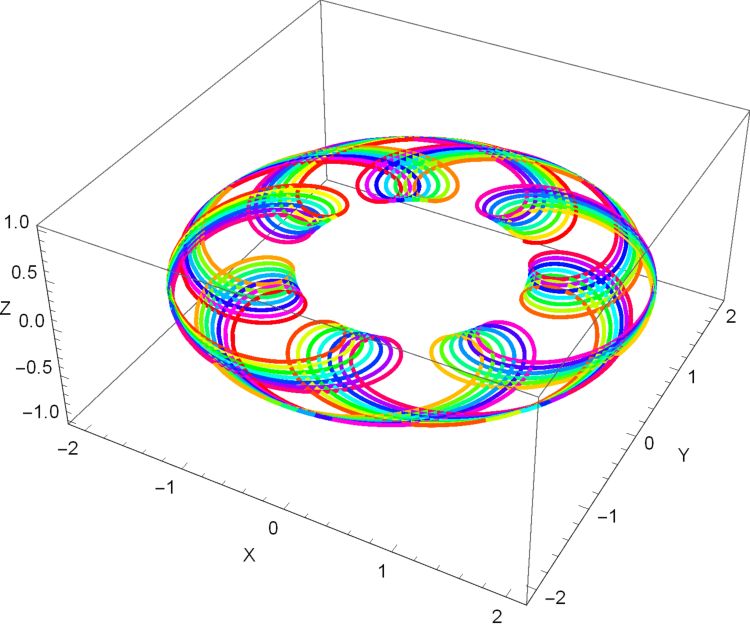}\\
	\includegraphics[scale=0.58]{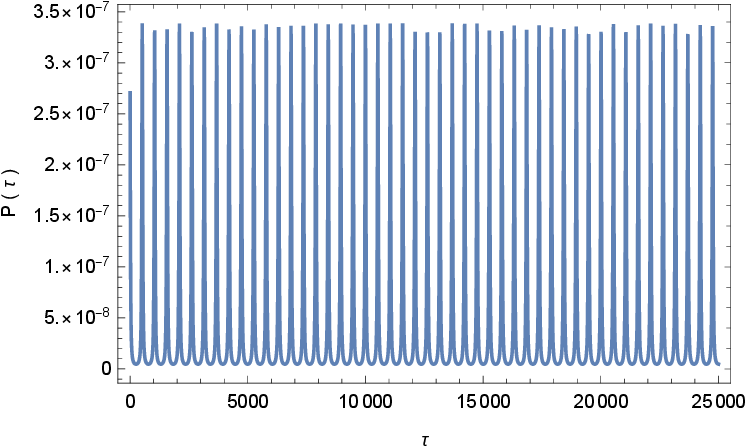}\qquad \qquad
	\includegraphics[scale=0.58]{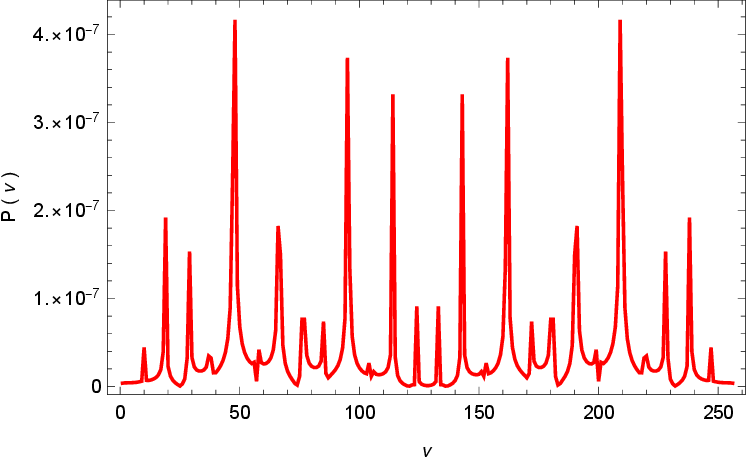}
	\caption{Trajectory, radiation power, and radiation spectrum for the Classical Relativistic St\"{o}rmer Problem (CRSP) for $\vec{R}_0=(0.7,0.8,0)$, $\left|\vec{R}_0\right| =1.063$,  $\vec{V}_0=(0.010,0,0)$, $V_0=\left|\vec{V}_0\right|=0.01$, and $\gamma =16$. \label{Fig4}}
\end{figure*}

A first example of relativistic motion, with $\gamma = 16$ and the initial velocity oriented purely along the $X$ axis, is presented in Fig.~\ref{Fig4}.

Qualitatively, the trajectory resembles the non-relativistic case: it is regular, periodic, yet displays a complex three-dimensional structure. The particle remains trapped inside the magnetic field and traces out a distinctive global pattern. The emitted electromagnetic power has constant amplitude, but its peak values are notably lower than in the non-relativistic regime; for the configuration of Fig.~\ref{Fig4} this reduction is primarily due to the smaller value of $V_0$. The radiation spectrum is similar to the non-relativistic one, with several well-defined peaks, but the overall spectral amplitude is significantly decreased.

\begin{figure*}[!htbp]
	\includegraphics[scale=0.54]{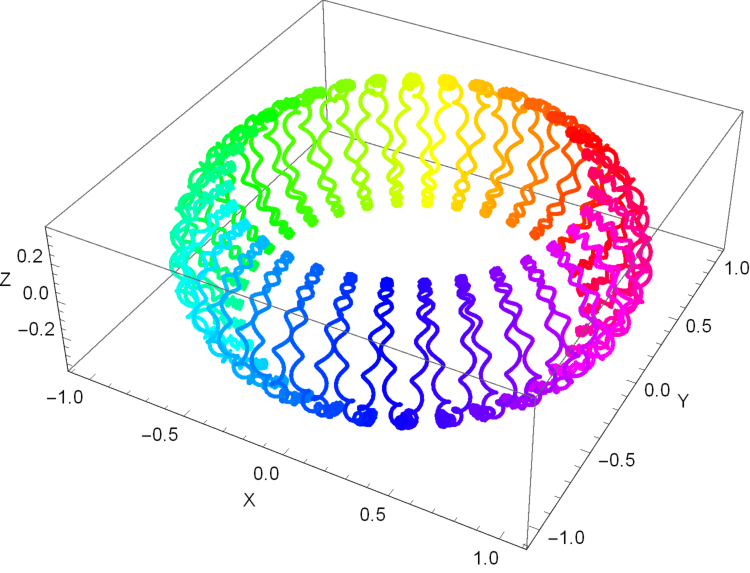}\\
	\includegraphics[scale=0.58]{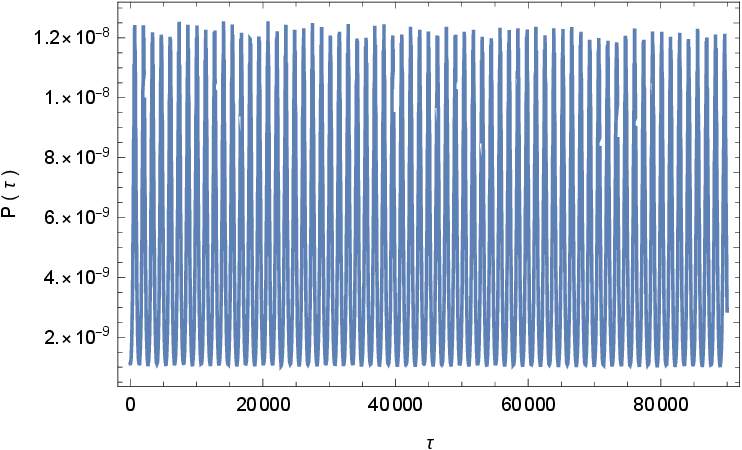}\qquad \qquad
	\includegraphics[scale=0.58]{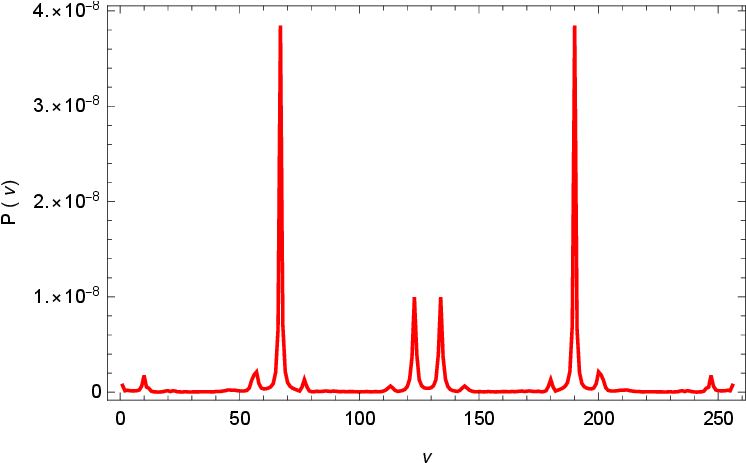}
	\caption{Trajectory, radiation power, and radiation spectrum for the Classical Relativistic St\"{o}rmer Problem (CRSP) for $\vec{R}_0=(0.7,0.8,0)$, $\left|\vec{R}_0\right| =1.063$,  $\vec{V}_0=(10^{-4},10^{-3},10^{-3})$, $V_0=\left|\vec{V}_0\right|=0.0014$, and $\gamma =25$. \label{Fig5}}
\end{figure*}

The case of a relativistic particle with all three components of the initial velocity non-zero is illustrated for $\gamma = 25$ in Fig.~\ref{Fig5}.

The trajectory remains confined and globally periodic, but its local structure is extremely intricate. It comprises a number of irregular three-dimensional motifs that are traversed over short time intervals; these local patterns are then glued together into a stable, globally periodic orbit. The emitted power is again periodic with diminished amplitude. The radiation spectrum is dominated by two prominent peaks, indicating sharp maxima in the spectral energy distribution.

\begin{figure*}[!htbp]
	\includegraphics[scale=0.54]{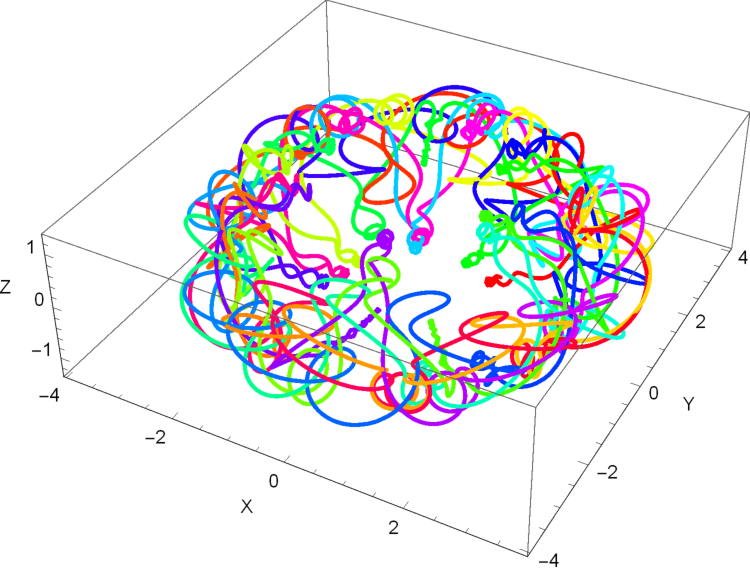}\\
	\includegraphics[scale=0.58]{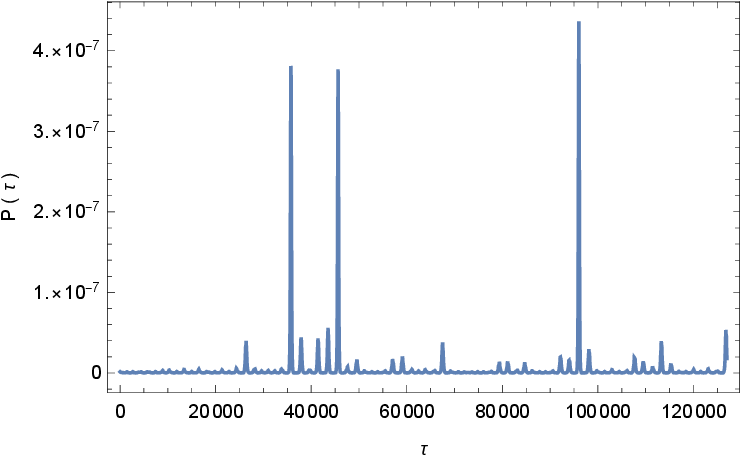}\qquad \qquad
	\includegraphics[scale=0.58]{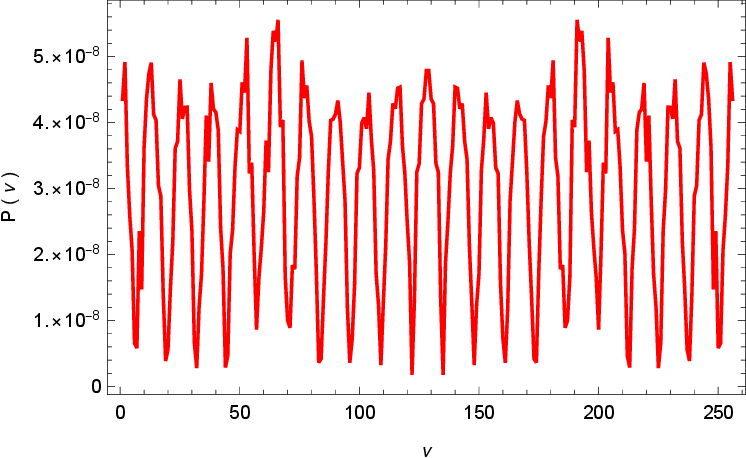}
	\caption{Trajectory, radiation power, and radiation spectrum for the Classical Relativistic St\"{o}rmer Problem (CRSP) for $\vec{R}_0=(1.9,1.9,0.9)$, $\left|\vec{R}_0\right| =2.833$,  $\vec{V}_0=(10^{-3},2\times 10^{-3},3\times 10^{-3})$, $V_0=\left|\vec{V}_0\right|=0.0037$, and $\gamma =5$. \label{Fig6}}
\end{figure*}

Finally, we consider a chaotic relativistic motion obtained for arbitrary initial conditions and $\gamma = 5$, displayed in Fig.~\ref{Fig6}.

The particle is still confined by the dipole field, but its trajectory is highly irregular and shows no trace of periodicity. As time advances new features continuously appear, signalling an increasingly complex evolution. The radiated power time series differs markedly from the previous cases: it consists of distinct emission spikes rising above a relatively constant background, rather than a clean periodic signal. These bursts arise from sudden variations in the particle's acceleration and do not obey any regular temporal sequence. The overall evolution is strongly dependent on the initial conditions. In contrast, the power spectrum is remarkably smooth, exhibiting only very small amplitude variations and no pronounced spectral peaks. This suggests that, despite the chaotic character of the motion, the radiation is distributed rather uniformly across the accessible frequencies.

\section{Discussions and final remarks}\label{sect4}

In this work we have extended the classical St\"{o}rmer problem to the relativistic regime, deriving the equation of motion of a charged particle in a dipole magnetic field in a fully covariant form. The relativistic correction enters solely through the constant Lorentz factor \(\gamma\), which is tied to the conserved particle energy. The underlying dipole magnetic field is given in Cartesian coordinates by
\begin{equation}
	\vec{B} = -\frac{B_{0}R_{0}^{3}}{r^{5}}\Bigl(3xz\,\vec{i} + 3yz\,\vec{j} + (2z^{2}-x^{2}-y^{2})\,\vec{k}\Bigr),
\end{equation}
where \(\vec{i},\vec{j},\vec{k}\) are the unit vectors along the coordinate axes.

Owing to their strongly nonlinear structure, the equations of motion in a dipole field rarely admit exact analytical solutions. Nevertheless, we have derived a closed parametric solution for the planar case, which generalises the simple harmonic circular orbit by allowing a time-dependent amplitude and frequency. In the ultra-relativistic limit this solution reduces to an explicit elementary form, from which the radial growth and angular advance can be read off directly.

In the extreme relativistic regime, when the Lorentz factor is large enough that the terms \(\gamma R^{3} \gg V_{X}, V_{Y}\), \(\gamma R^{5} \gg Z(Z V_{Y} - Y V_{Z})\), and analogous combinations hold, the equations of motion simplify dramatically:
\begin{equation}
	\frac{d^{2}X}{d\tau^{2}} \approx 0,\quad \frac{d^{2}Y}{d\tau^{2}} \approx 0,\quad \frac{d^{2}Z}{d\tau^{2}} \approx 0,
\end{equation}
with the solution
\begin{eqnarray*}
	X(\tau) = X_{0} + V_{0X}\tau,&&\quad Y(\tau) = Y_{0} + V_{0Y}\tau,
		\nonumber \\
	Z(\tau) &=& Z_{0} + V_{0Z}\tau.
\end{eqnarray*}
Thus the particle moves along a straight line and, for sufficiently high \(\gamma\), escapes to infinity. A similar escape occurs if the velocity components satisfy the alignment conditions
\begin{eqnarray*}
	\frac{V_{X}}{V_{Y}} = \frac{X}{Y},\qquad && \frac{V_{Z}}{V_{X}} = \frac{1}{X}\!\left(Z-\frac{R^{2}}{3Z}\right),
		\nonumber \\ 
	\frac{V_{Z}}{V_{Y}} &=& \frac{1}{Y}\!\left(Z-\frac{R^{2}}{3Z}\right),
\end{eqnarray*}
in which case the magnetic force identically vanishes and the trajectory remains rectilinear.

To probe the stability of planar orbits we performed a linear perturbation analysis around a fixed equatorial position \((X_{0},Y_{0})\). Writing \(X = X_{0}+\delta X\), \(Y = Y_{0}+\delta Y\), the perturbed equations become
\begin{equation}
	\frac{d^{2}\delta X}{d\tau^{2}} = -\frac{1}{\gamma R_{0}^{3}}\frac{d\delta Y}{d\tau},\qquad
	\frac{d^{2}\delta Y}{d\tau^{2}} = \frac{1}{\gamma R_{0}^{3}}\frac{d\delta X}{d\tau},
\end{equation}
with \(R_{0}^{2}=X_{0}^{2}+Y_{0}^{2}\). Combining them yields a third-order equation for \(\delta X\),
\begin{equation}\label{89}
	\frac{d^{3}\delta X}{d\tau^{3}} + \frac{1}{\gamma^{2} R_{0}^{6}}\frac{d\delta X}{d\tau} = 0,
\end{equation}
whose solution for the initial data \(\delta X(0)=0\), \(\delta X'(0)=\delta V_{0X}\), \(\delta X''(0)=\delta A_{0X}\) is
\begin{equation}
	\delta X(\tau) = \gamma R_{0}^{3}\Bigl[ \delta V_{0X} \sin\!\frac{\tau}{\gamma R_{0}^{3}} + \gamma R_{0}^{3} \delta A_{0X}\Bigl(1-\cos\!\frac{\tau}{\gamma R_{0}^{3}}\Bigr) \Bigr].
\end{equation}
The \(Y\) perturbation obeys
\begin{equation}
	\frac{d^{2}\delta Y}{d\tau^{2}} - A_{0X}\sin\!\frac{\tau}{\gamma R_{0}^{3}} - \frac{V_{0X}}{\gamma R_{0}^{3}}\cos\!\frac{\tau}{\gamma R_{0}^{3}} = 0,
\end{equation}
with the general solution
\begin{eqnarray}
	\delta Y(\tau) &=& \gamma R_{0}^{3} V_{0X} + \bigl(\gamma A_{0X} R_{0}^{3} + V_{0Y}\bigr)\tau \nonumber\\
	&& - \gamma R_{0}^{3}\Bigl( \gamma A_{0X} R_{0}^{3} \sin\!\frac{\tau}{\gamma R_{0}^{3}} + V_{0} \cos\!\frac{\tau}{\gamma R_{0}^{3}} \Bigr),
\end{eqnarray}
where \(\delta Y(0)=0\) and \(\delta Y'(0)=V_{0Y}\). While the \(X\) perturbation is purely oscillatory, the \(Y\) perturbation contains a secular term that grows linearly with time, indicating generic instability. However, for perturbations satisfying \(\gamma A_{0X} R_{0}^{3} + V_{0Y} = 0\) the secular term vanishes, and the motion remains bounded and stable. This fine-tuned stability condition may be relevant for the long-term confinement of particles in radiation belts.

We have also performed a systematic analysis of the electromagnetic radiation emitted by the particle. The emitted power and its Fourier spectrum were computed for both regular and chaotic trajectories, revealing distinct spectral signatures that reflect the underlying dynamics.

The St\"{o}rmer problem, in all its versions, inevitably rests on a number of physical simplifications: the field is taken to be a pure dipole, radiative and dissipative effects are neglected, and gravitation is omitted. Despite these idealisations, it remains a remarkably effective framework for capturing the essential dynamics of charged particles in magnetised environments. The relativistic extension developed here provides a richer phase space, admitting periodic, quasi-periodic, and chaotic orbits alongside escape trajectories.

Future work could build on the present study in several directions. Including the radiation reaction self-consistently would allow one to follow the gradual energy loss and the eventual fate of the trapped particles. Extending the analysis to more realistic magnetic field configurations that incorporate higher multipole moments, or to curved spacetime backgrounds, would bring the model closer to the conditions encountered around neutron stars and black holes. These generalisations would deepen our understanding of the interplay between relativistic dynamics, magnetic confinement, and radiative processes in compact astrophysical objects.

%%%%%%%%%%%%%%%%%%%%%%%%%%%%%%%%%%%%%%%%%%%%%%%%%%%%%%%%%%%%%
\acknowledgments{
FSNL acknowledges funding from the Funda\c{c}\~{a}o para a Ci\^{e}ncia e a Tecnologia (FCT) through national funds under the research grant UID/04434/2025 (DOI 10.54499/UID/04434/2025), and support from the FCT Scientific Employment Stimulus contract with reference CEECINST/00032/2018.}
%%%%%%%%%%%%%%%%%%%%%%%%%%%%%%%%%%%%%%%%%%%%%%%%%%%%%%%%%%%%%


\begin{thebibliography}{99}


\bibitem{St1} C. St\"{o}rmer, \textit{Arch. Sci. Phys. Nat.} \textbf{1907},
24, 5.

\bibitem{St2} C. St\"{o}rmer, \textit{Arch. Sci. Phys. Nat.} \textbf{1907},
24, 113.

\bibitem{St3} C. St\"{o}rmer, \textit{Arch. Sci. Phys. Nat.} \textbf{1907},
24, 221.

\bibitem{St4} C. St\"{o}rmer, \textit{Astrophysical Journal} \textbf{1913},
38, 311.

\bibitem{St4a} C. St\"{o}rmer, \textit{Geofys\'{\i}sk Publikationer} \textbf{%
1921}, 1, 269.

\bibitem{St5} C. St\"{o}rmer, \textit{Terrestrial Magnetism and Atmospheric
Electricity} \textbf{1917}, 22, 97.

\bibitem{St6} C. St\"{o}rmer, \textit{Astrophysica Norvegica} \textbf{1934},
1, 1.

\bibitem{St7} C. St\"{o}rmer, \textit{The Polar Aurora}, Clarendon Press,
Oxford, UK, \textbf{1955}.

\bibitem{B1} A. Dragt, \textit{Rev. Geophys.} \textbf{1965}, 3, 255.

\bibitem{B2} A. Dragt and J. M. Finn, \textit{J. Geophys. Res.} \textbf{1976}, 81, 2327.

\bibitem{B3} M. Walt, Introduction to Geomagnetically Trapped Radiation, Cambridge Atmospheric and Space Science Series, Cambridge University Press, 1994

\bibitem{Jack} J. D. Jackson, Classical Electrodynamics,  John Wiley \& Sons, Hoboken, NJ, USA, 1999 

\bibitem{LL} L. D. Landau and E. M. Lifshitz, The classical theory of fields, Pergamon Press, Oxford, UK, 1994

\bibitem{Int} M. A. Almeida, I. C. Moreira, and H. Yoshida, \textit{J. Phys.
A Math. Gen.} \textbf{1992}, 25, L227.

\bibitem{Dilao} R. Dil$\tilde{{\rm a}}$o and R. Alves-Pires (2007), Chaos in the St\"{o}rmer Problem. In: Staicu, V. (eds) Differential Equations, Chaos and Variational Problems. Progress in Nonlinear Differential Equations and Their Applications, vol 75. Birkh\"{a}user, Basel, pp 175-194.

\bibitem{VA1} Y. Y. Shprits et al., Nature Physics \textbf{9}, 699-703
(2013).

\bibitem{VA2} J. H. Zhang, L. Y. Li, Y. W. Yao, K. X. Cheng, and L. Yang,
Journal of High Energy Astrophysics 52, 100568 (2026).

 \bibitem{Schust}   R. Schuster and K. O. Thielheim, {\it J. Phys. A: Math. Gen.} {\bf 1987}, 20, 5511.
   
    \bibitem{How}  J. E. Howard, M. Hor\'{a}nyi, and G. R.  Stewart,  {\it Phys. Rev. Lett.} {\bf 1999}, 83, 3993.
    
    \bibitem{Dull} H. R. Dullin,  M. Hor\'{a}nyi,  J. E.  Howard, {\it Physica D} {\bf 2002},  171, 178.
    
     \bibitem{In} I$\tilde{{\rm n}}$arrea M. et al.,  {\it Physica D} {\bf 2004},  197, 242.

\bibitem{In1} I$\tilde{{\rm n}}$arrea M. et al., {\it Chaos, Solitons and Fractals} {\bf 2009}, 42, 155.

\bibitem{Epp0} V. Epp,  M. A.  Masterova,  {\it Astrophysics and Space Science} {\bf 2014}, 353, 473.

\bibitem{Epp} V. Epp,  O. N. Pervukhina, {\it Monthly Notices of the Royal Astronomical Society} {\bf 2018},  474, 5330.

\bibitem{Hal} V. V. Markellos and A. A. Halioulias, {\it Astrophysics and Space Science} {\bf 1977},  51,  177. 

\bibitem{Mark} V. V. Markellos and C. Zagouras, {\it Astronomy and Astrophysics} {\bf 1977}, 61, 505-514.

\bibitem{Ozturk} M. K. \"{O}zt\"{u}rk, {\it American Journal of Physics} {\bf 2012},  80(5), 420.


  \bibitem{Pina} E. Pina, E.  Cort\'{e}s,  {\it European Journal of Physics} {\bf 2016}, 37,  065009.
  
  \bibitem{Kol} E. K. Kolesnikov, {\it Geomagnetism and Aeronomy} {\bf 2017}, 57,  137.
  
  \bibitem{Leg} A. Leghmouche, N. Mebarki,  A. Benslama,  {\it New Astronomy} {\bf 2023}, 98,  101931.
  
  \bibitem{Ersh} S. Ershkov,  E. Prosviryakov, D. Leshchenko,  N. Burmasheva,   
{\it Mathematical Methods in the Applied Sciences} {\bf 2023}, 46,  19364.

\bibitem{Asadi} M. Asadi-Zeydabadi, C. S.  Zaidins  {\it Results in Physics} {\bf 2019}, 12, 2213. 

%\bibitem{Sfarti} A. Sfarti, {\it Journal of Progress in Engineering
%and Physical Science} {\bf 2023}, 

\bibitem{Ersh1} S. V. Ershkov,  {\it J. Appl. Comput. Mech.} {\bf 2026},  12, 31.

\bibitem{Moc} T. Harko and G. R. Mocanu, {\it Annalen der Physik} {\bf 2025}, 537 , e00415.

\bibitem{Lyut} M. V. Barkov and M. Lyutikov, arXiv:2506.20515 [astro-ph.HE]
(2025).

\bibitem{Th} K. S. Thorne, \textit{Astrophysical Journal Supplement} {\bf 1963}, 8, 1. 

\bibitem{Pap} D. B. Papadopoulos, I. Contopoulos, K. D. Kokkotas, N. Stergioulas, \textit{General Relativity and Gravitation} {\bf 2015}, 47, 49. 

\bibitem{Bur} T. M. Burinskaya, M. M. Shevelev, \textit{Plasma Physics Reports} {\bf 2016}, 42, 929. 

\bibitem{Bur1}  T. M. Burinskaya, M. M. Shevelev, \textit{Plasma Physics Reports} {\bf 2017}, 43, 910.

\end{thebibliography}
\end{document}